\documentclass[preprints,article,accept,moreauthors,pdftex]{Definitions/mdpi} 

\firstpage{1} 
\pubvolume{xx}
\issuenum{1}
\articlenumber{5}
\pubyear{2019}
\copyrightyear{2019}
\history{Received: date; Accepted: date; Published: date}

\Title{Wind profiling in the lower atmosphere from wind-induced perturbations to multirotor UAS}

\Author{Javier Gonz\'alez-Rocha$^{1}*$, Stephan F.~J. De Wekker $^{2}$, Shane D. Ross$^{1}$ and Craig A. Woolsey $^{1,}$}

\AuthorNames{Javier Gonz\'alez-Rocha, Firstname Lastname and Firstname Lastname}

\address{%
$^{1}$ \quad Department of Aerospace and Ocean Engineering, Virginia Tech, Blacksburg, VA 24060, U.S.A; javig86@vt.edu (J.G.R.); sdross@vt.edu (S.D.R); cwoolsey@vt.edu (C.A.W.)\\
$^{2}$ \quad Department of Environmental Sciences, University of Virginia, Charlottesville, VA 22903, U.S.A.;
dewekker@virginia.edu (S.F.J.D)}

\corres{Correspondence: e-mail: javig86@vt.edu; Tel.:+1-540-231-2019}

\abstract{We present a model-based approach to wind velocity profiling using motion perturbations of a multirotor unmanned aircraft system (UAS) in both hovering and steady ascending flight. A state estimation framework was adapted to a set of closed-loop rigid body models identified for an off-the-shelf quadrotor. The quadrotor models used for wind estimation were characterized for hovering and steady ascending flight conditions ranging between 0 and 2 m/s. The closed-loop models were obtained using system identification algorithms to determine model structures and estimate model parameters. The wind measurement method was validated experimentally above the Virginia Tech Kentland Experimental Aircraft Systems Laboratory by comparing quadrotor and independent sensor measurements from a sonic anemometer and two SoDARs. Comparison results demonstrated quadrotor wind estimation in close agreement with the independent wind velocity measurements. Wind velocity profiles were difficult to validate using time-synchronized SoDAR measurements, however. Analysis of the  noise intensity and signal-to-noise ratio of the SoDARs proved that close-proximity quadrotor operations can corrupt wind measurement from SoDARs.}

\keyword{Unmanned Aircraft Systems, System Identification, Wind Estimation, Multi-Rotor, Drone, Atmospheric Science, Wind Profile, Boundary Layer Meteorology}

\usepackage{pmat}
\usepackage{booktabs} 
\usepackage[normal]{subfigure}
\usepackage{booktabs}
\usepackage{multirow}
\usepackage{amsfonts}
\usepackage{amssymb}

\newcommand{\beq}{\begin{equation}}
\newcommand{\eeq}{\end{equation}}
\newcommand{\overliner}{\begin{array}}
\newcommand{\earr}{\end{array}}
\newcommand{\beqarr}{\begin{eqnarray}}
\newcommand{\eeqarr}{\end{eqnarray}}
\newcommand{\beqar}{\begin{eqnarray*}}
\newcommand{\eeqar}{\end{eqnarray*}}
\newcommand{\bef}{\begin{figure}}
\newcommand{\eef}{\end{figure}}
\newcommand{\reff}[1]{(\ref{#1})}

\newcommand{\bm}[1]{\mbox{\boldmath$#1$}}

\begin{document}
\section{Introduction}
\label{s:introduction}
Measuring wind velocity near the Earth's surface is critical to understanding the surface-atmosphere interactions driving the dynamic state of the atmospheric boundary layer (ABL). How the ABL evolves with space and time influences phenomena that impact public health and safety~\cite{gonzalez2019sensing,barbieri2019intercomparison,jacob2018considerations,chilson2019moving,smith2017catalyzing}. For example, the transport of air pollutants, pollen and spores~\cite{villa2016overview,nolan2018coordinated,nolan2019method,carranza2018vista}, wind power supply to smart grid systems~\cite{chao2010surface,Fairley2018building,phuangpornpitak2013opportunities,colak2015critical,wildmann2017measuring}, forecast of local weather~\cite{barbieri2019intercomparison,jacob2018considerations,chilson2019moving,smith2017catalyzing}, air traffic control at airports ~\cite{alsalous2017evaluation,tang2010accurate,tang2011lagrangian,knutson2015lagrangian}, the spread and management of wildfires~\cite{rabinovich2018toward,da2017unmanned,al2017review,Xingetal2019}, and emissions mitigation of greenhouse gases~\cite{duren2019california,smith2017fugitive,andersen2018auav,roldan2015mini} are all affected by the dynamic state of the ABL. Therefore, mitigation of adverse conditions affected by the dynamic state of the ABL requires accurate measurements of wind velocity over micro- and mesoscale domains~\cite{barbieri2019intercomparison,greene2019environmental,varentsov2019experience}. However, observations of wind velocity at high spatial resolution are difficult due to the cost and limited mobility of conventional atmospheric sensing technology.  

Advancing capabilities for wind sensing with multirotor unmanned aircraft systems (UAS) can help fill the existing gap in ABL observations~\cite{jacob2018considerations,chilson2019moving,smith2017catalyzing}. This is due to the effectiveness of multirotor UAS for probing the ABL over complex terrain or water where reliable operation of in situ or remote sensors is prohibitively difficult or difficult. In general, multirotor UAS are mobile, portable, low cost, and easy to operate. Improving upon existing wind sensing algorithms to expand the flight envelope in which a multirotor can accurately measure the wind velocity can significantly enhance their utility for on-demand targeted observations inside the ABL. Increasing the capability of multirotor UAS for wind sensing in this way can supplement ground-based and airborne atmospheric observations currently used to characterize the ABL.

Existing wind sensing approaches with multirotor UAS consist of direct and indirect methods. Direct methods involve the retrieval of wind velocity from a flow sensor onboard a multirotor. Examples include various types of anemometers~\cite{wolf2017wind,de2014designing,donnell2018wind,hollenbeck2018wind,Hollerbeck2019pitch} and other air data systems~\cite{prudden2016flying}. The choice of sensor depends on the sensor size and power requirements, the aircraft payload capacity, and the airframe configuration of the multirotor aircraft. Indirect methods, on the other hand, estimate wind velocity from wind induced perturbations to the aircraft motion and do not require a separate airflow sensor. Conventional model-based approaches to wind estimation have involved kinematic~\cite{neumann2015real,brosy2017simultaneous}, point mass~\cite{palomaki2017wind,gonzalez2019sensing,donnell2018wind}, and rigid body models~\cite{gonzalez2019sensing} of control-augmented quadrotor dynamics, which characterize how a quadrotor responds to disturbances under feedback stabilization. A comparison of all three models in~\cite{gonzalez2019sensing} demonstrated that both the accuracy and bandwidth of wind estimates increases with fidelity of the vehicle motion model.

To date, model-based wind estimation approaches have only incorporated models that are appropriate for \textit{hovering} flight~\cite{gonzalez2019sensing}. Measuring wind velocity only while hovering limits the speed at which the aircraft can sample the lower atmosphere. The limitation of stationary sampling is largely due to the limited endurance of multirotor aircraft (typically less than 20 minutes). However, many research and operational applications require atmospheric sampling over horizontal and vertical distances. Therefore, there is a need to develop wind estimation algorithms that allow a multirotor UAS to move while accurately measuring wind velocity within the ABL. 

This paper presents a method for estimating vertical profiles of the horizontal wind velocity using a dynamic rigid body model of a quadrotor in hovering and steady-ascending equilibrium flight conditions. The method presented here, referred to as the 
{\it dynamic rigid body wind profiling} method or
{\it DRBWindPro} method for short,  is an extension of the wind sensing algorithm presented in \cite{gonzalez2019sensing} to measure wind velocity in hovering flight. The extension of the wind sensing algorithm incorporates dynamic rigid body models characterized from system identification for equilibrium flight conditions corresponding to steady ascent rates ranging from 0 to 2~m/s. The models from system identification were used to estimate the wind velocity in the vicinity of ground-based in situ and remote atmospheric sensors. Quadrotor wind estimates and wind measurements from ground-based atmospheric sensors were then compared to determine the accuracy of the DRBWindPro method. 

The organization of this paper is as follows. Section~\ref{sec:materials_and_methods} introduces materials and methods used for model-based wind estimation. This section includes the formulation of aircraft dynamics, system identification of aircraft models, and the design of a state observer for wind estimation. The ground-based wind measurement methods are described in Section~\ref{sec:experimental_validation_of_wind_estimates}. In Section~\ref{sec:results} results from system identification experiments and comparison of multirotor wind velocity measurements with ground-based measurements are presented. Section~\ref{sec:discussion} presents a thorough discussion of results from system identification and from comparing multirotor and ground-based wind measurements. Finally, a summary of findings and future work to extend the utility of multirotor UAS for wind sensing are presented in Section~\ref{sec:conclusion}.

\section{Materials and Methods}
\label{sec:materials_and_methods}

\subsection{Modeling Framework}
\label{ss:modeling_framework}
The equations of motion for a control-augmented (i.e., feedback-stabilized) quadrotor can be expressed as a system of first-order, nonlinear, time-invariant ordinary differential equations~\cite{gonzalez2017measuring}:
\beq \bm{\dot{x}} = \bm{f}(\bm{x},\bm{u},\bm{w}(t,\bm{x})),\hspace{1cm} \bm{x}(t_0) = \bm{x}_0\eeq
 relating the rate of change $\bm{\dot{x}}$ of the vehicle's 12-dimensional state $\bm{x}$ (i.e., position, attitude, velocity, and angular velocity), to the state itself, the control inputs $\bm{u}$, and wind disturbances $\bm{w}(t,\bm{x})$ varying over time and space. Moreover, when the aircraft motion is modeled as a small perturbation from some equilibrium flight condition that corresponds to a constant vertical ascent speed denoted by $V_{z_\mathrm{eq}}$, the nonlinear dynamics describing the control-augmented motion of the quadrotor is well approximated by a linear model. As a result, one may infer wind velocity from wind-induced motion perturbations to a quadrotor employing estimation theory developed for linear systems.

 Linear approximations of quadrotor dynamics for wind estimation are considered in this study for hovering and steady-ascending motions satisfying trim flight conditions. For a quadrotor, trim flight conditions are satisfied when both translational rates $\bm{v}$ and rotational rates $\bm{\omega}$ remain constant over time, i.e., $\bm{\dot{v}}\equiv \bm{0}$ and $\bm{\dot{\omega}}\equiv \bm{0}$. Linear approximations of quadrotor dynamics for hovering and ascending flight are in the form,
 \beq \frac{d}{dt}\bm{{\tilde{x}}} = \bm{A}\bm{\tilde{x}}+ \bm{B}\bm{\tilde{u}}+\bm{\Gamma}\bm{w},\label{eqn:linearized_dynamics}\eeq
 where the vectors $\bm{\tilde{x}} = \bm{x}-\bm{x}_{\rm eq}$ and $\bm{\tilde{u}} = \bm{u}-\bm{u}_{\rm eq}$ denote, respectively, small deviations in the state and input vectors from their steady-state values. Additionally, the state matrix $\bm{A}\in\mathbb{R}^{12\times12}$ models unforced dynamics, the input matrix $\bm{B}\in\mathbb{R}^{12\times4}$ characterizes applied forcing, and the disturbance matrix $\bm{\Gamma}\in \mathbb{R}^{12\times3}$ captures wind-induced perturbations. This model form is used to sense wind velocity at different steady motion conditions (i.e., different steady ascent rates $V_{z_{\rm eq}}$) using state estimation once state, input, and disturbance matrices are known. 
 
 \subsection{Aircraft System Identification}
 \label{ss:aircraft_system_identification}
 
\begin{figure}[tbh!]
    \centering
    \includegraphics[width = 11cm]{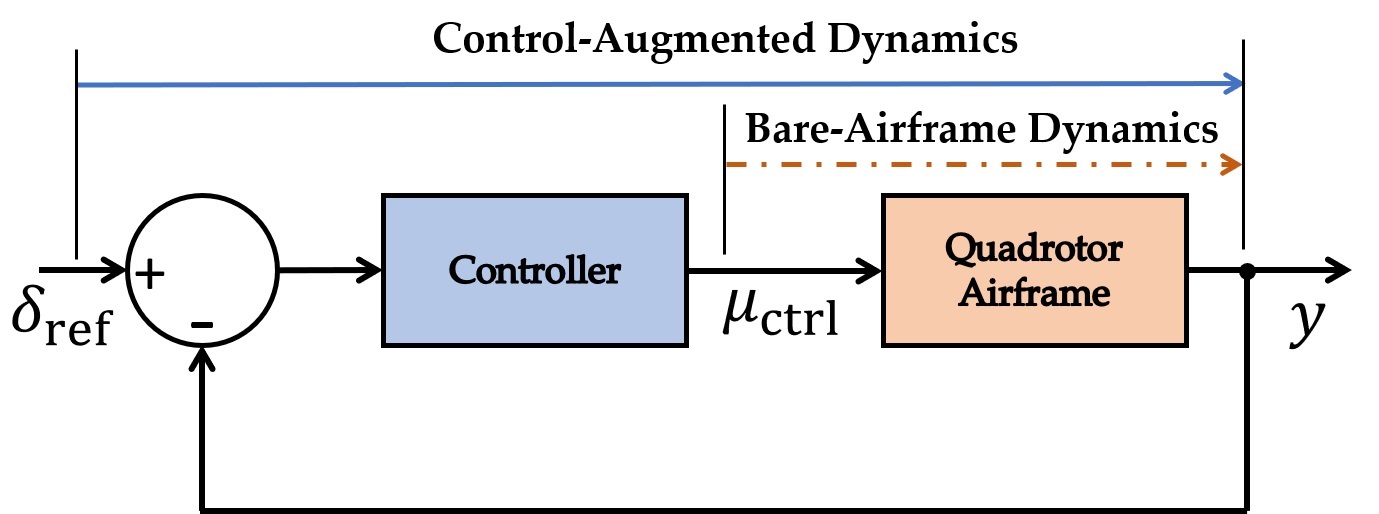}
    \caption{A schematic of input-output signals for closed-loop and open-loop mappings. }
    \label{fig:closed-loop_open-loop}
\end{figure}

 Aircraft system identification is used to characterize the state and input matrices $\bm{A}$ and $\bm{B}$ for a quadrotor flying in still air conditions (i.e., $\bm{w}(\bm{x},t) \approx \bm{0}$ m/s). In general, this modeling approach is a multi-faceted process that relies on input-output flight test data to characterize bare-airframe or control-augmented dynamic models for an aircraft, depending on application. Figure~\ref{fig:closed-loop_open-loop} shows a schematic of the inputs $\bm{u}$ and outputs $\bm{y}$ used to identify bare-airframe and control-augmented models. A bare-airframe model, assuming actuator dynamics to be negligible, is identified using control signals from the flight controller $\bm{\mu}_{\rm ctrl}$ and the vehicle's measured dynamic response $\bm{y}$. A control-augmented model, alternatively, is identified using the reference signal $\bm{\delta}_{\rm r}$ from pilot-induced joystick commands and the vehicle's measured dynamic response $\bm{y}$. Which model is identified depends on its application. For wind estimation purposes with an off-the-shelf quadrotor, we use the latter because it does not require knowledge of the onboard flight controller architecture. 
 
\begin{figure}[b]
    \centering
    \includegraphics[width = 13cm]{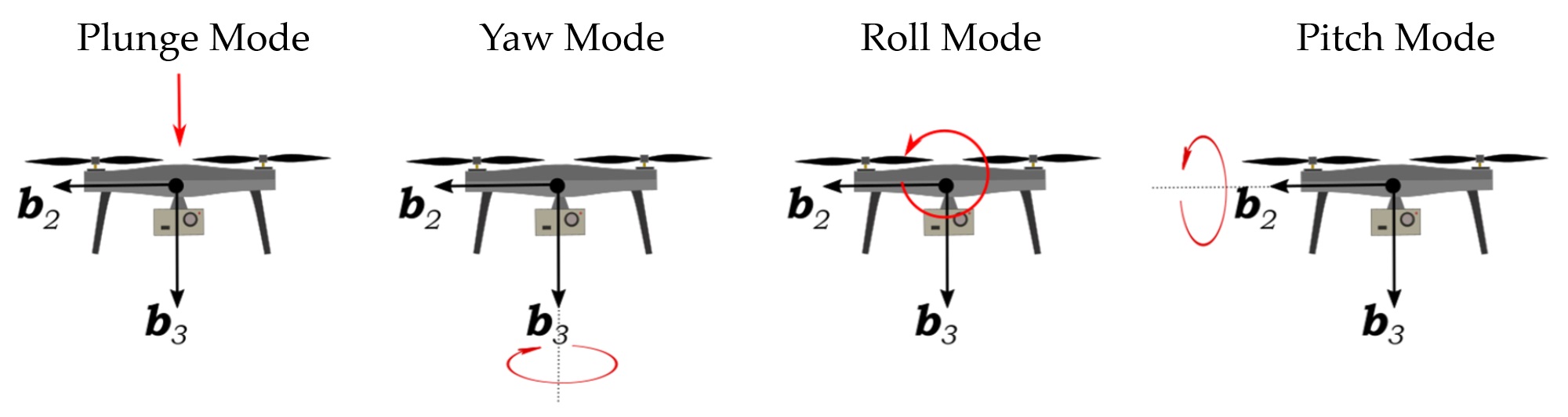}
    \caption{Quadrotor plunge, yaw, roll, and pitch modes }
    \label{fig:quadrotor_modes}
\end{figure}

The quadrotor models from system identification are for steady-state equilibrium flight conditions corresponding to the hovering and steady ascending flight: $V_{z_{\rm eq}} = \{0.0,0.5,1.0,1.5,2.0\}$~m/s. The identification of each model involved separately determining four sub-models that describe the plunge, yaw, roll, and pitch dynamics of the quadrotor; see Figure~\ref{fig:quadrotor_modes}. In this process, stepwise regression was used first to determine the parameter structure of each model. Results from stepwise regression were then used to estimate model parameters using an output error algorithm. This approach to system identification was used to minimize the set of parameters being estimated at one time and to avoid overparameterized models. 

\subsubsection{Multirotor UAS Platform}
\label{Multirotor_UAS_Platform}
\begin{figure}
    \centering
    \subfigure[Quadrotor platform]{\includegraphics[width=50mm]{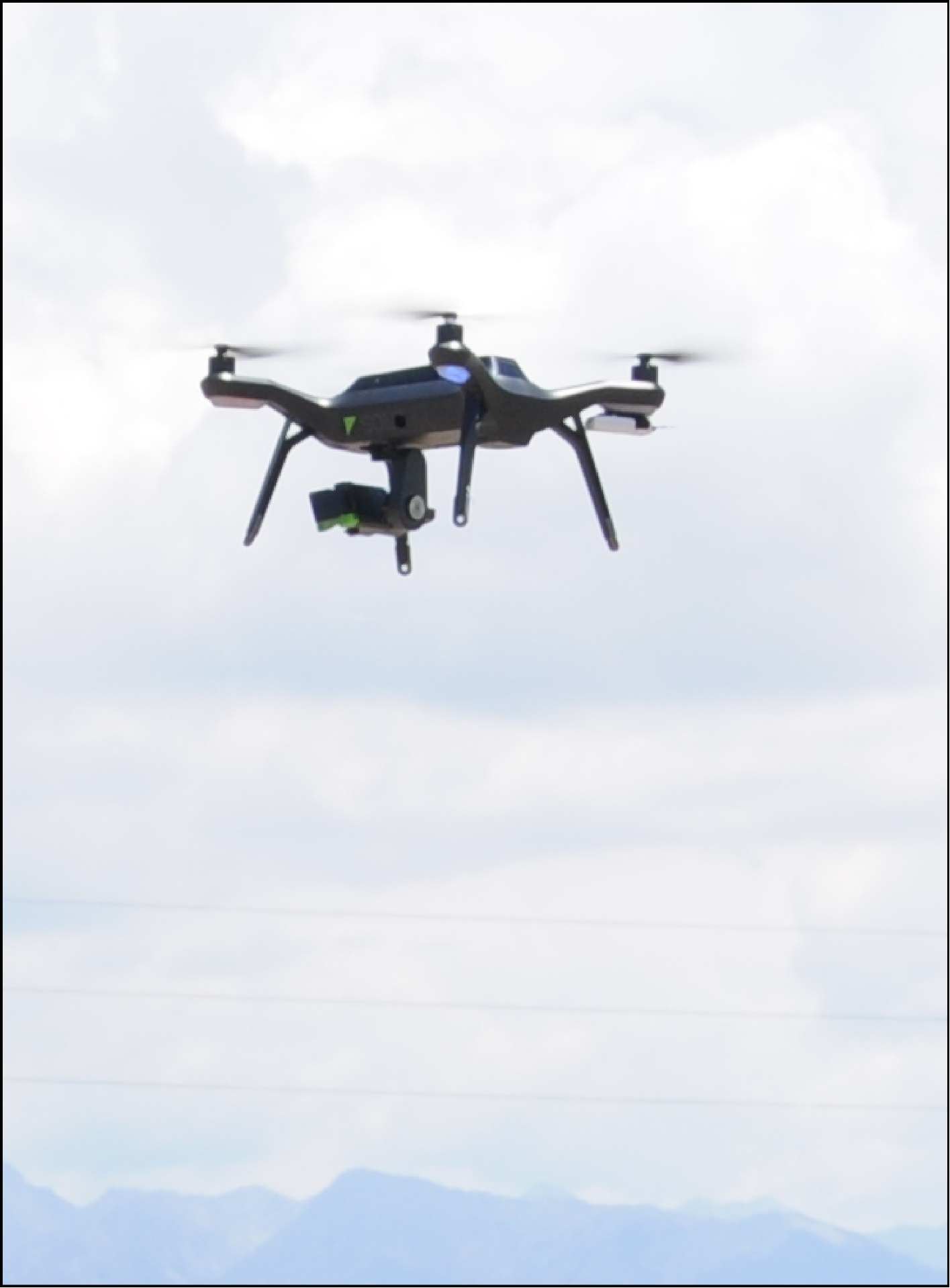}\label{subfig:sonic_anemometer}}\qquad
    \subfigure[Quadrotor dimensions]{\includegraphics[width=50mm]{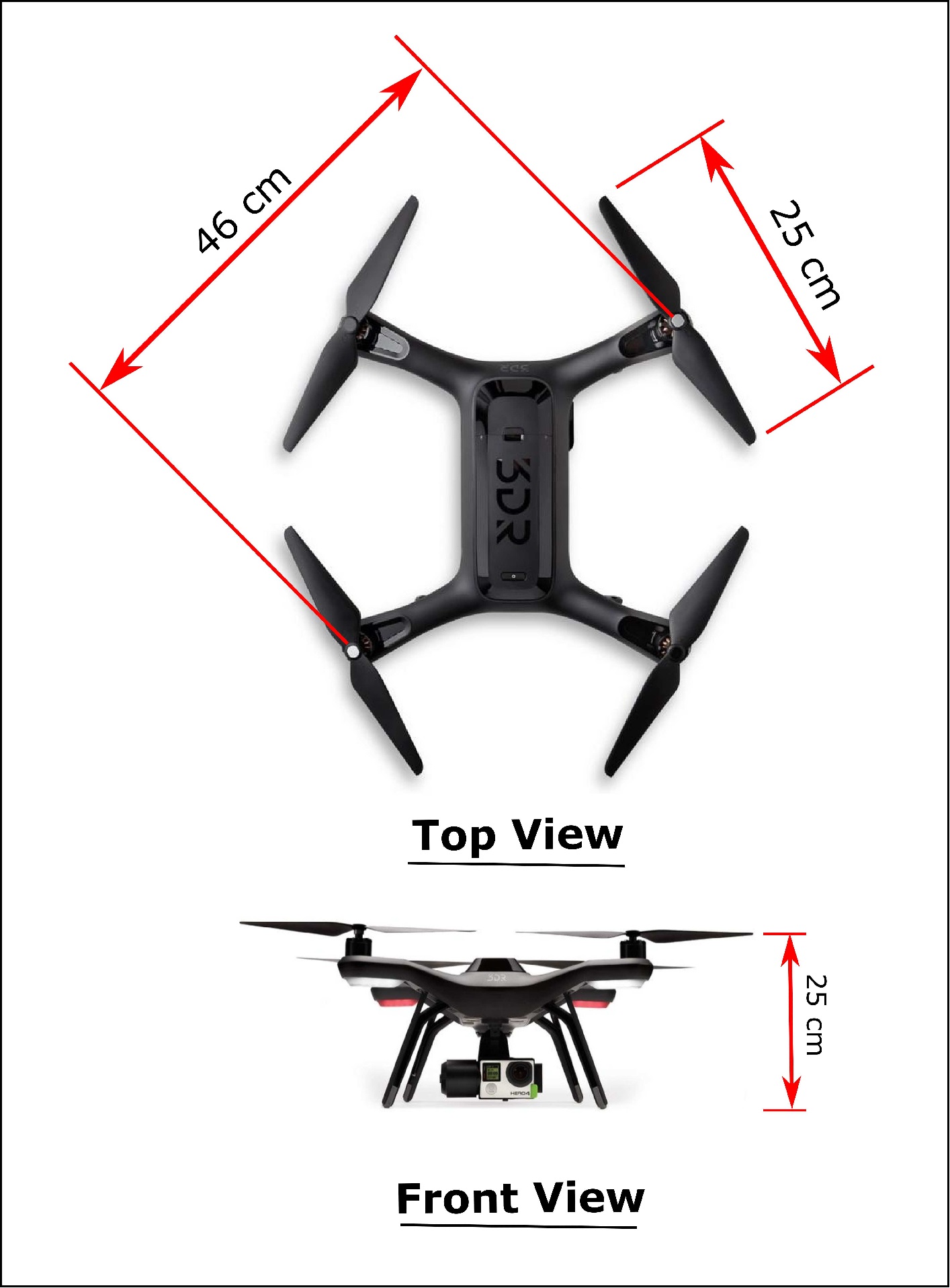}\label{subfig:SoDAR}}
    \caption{ Atmospheric sensors used for validation of UAS-based wind sensing. }
    \label{fig:3DR Solo}
\end{figure}

The multirotor UAS used to measure the wind velocity is an off-the-shelf 3DR Solo quadrotor shown in Figure~\ref{fig:3DR Solo}. This aircraft is 25 cm tall with a 46 cm diagonal between motor shafts.  Fully equipped with a lithium polymer battery pack and a 3-axis camera gimbal, the quadrotor weighs 1.5 kg and has a payload capacity of 0.5 kg. The propellers used with the quadrotor are a Master Airscrew $10\times 4.5$ propeller set. The quadrotor's autopilot is a Pixhawk 2.1 Green Cube manufactured by ProfiCNC. The autopilot operates using open-source Arducopter firmware and is compatible with MissionPlanner and Solex telemetry software. On board the Pixhawk 2.1 Green Cube are the sensors listed in Table \ref{table:pixhawk-integrated_sensors} that are part of the autopilot's attitude and heading reference system (AHRS).


\begin{table}[tbh!]
\centering
  \begin{tabular}{ccccccccccccccc}
    \toprule
    \multirow{2}{*}{State Measurement} & \multirow{2}{*}{Sate Variables} &&
      \multicolumn{4}{c}{ Sensor Type \& Sampling Rate }\\
      \cline{4-7} \
       &&& \multicolumn{2}{c}{ Direct }  & \multicolumn{2}{c}{ Indirect }    \\
      \toprule \midrule
\multirow{2}{*}{Position} & \multirow{2}{*}{ $ \{x,y,z\}$} & & \multirow{2}{*}{GPS} & \multirow{2}{*}{5Hz} &   Barometer & 8 Hz    \\ &&&&& Extended Kalman Filter & 8 Hz \\\midrule
\multirow{3}{*}{Attitude} &  \multirow{3}{*}{$\{\phi,\theta,\psi \}$}  &&\multirow{3}{*}{---}& \multirow{3}{*}{---} & Gyroscope & 18 Hz  \\&&&&& Accelerometer & 18 Hz \\&&&&& Extended Kalman Filter & 8 Hz \\\midrule
 Translational & \multirow{2}{*}{ $\{u,v,w \} $ }  && \multirow{2}{*}{GPS} & \multirow{2}{*}{5Hz} & Accelerometer & 18 Hz    \\Velocity&&&&& Extended Kalman Filter & 8 Hz \\\midrule
 Angular Velocity &  $\{p,q,r\} $   && Gyroscope & 18 Hz & --- & ---    \\\midrule
  \bottomrule
  \end{tabular}
  \caption{State measurements from autopilot's AHRS.}
  \label{table:pixhawk-integrated_sensors}
  \end{table}

\subsubsection{System Identification Flight Testing}
\label{sss:system_identification_flight_testing}

System identification flight experiments were conducted outdoors in an open field adjacent to the Virginia Tech Kentland Experimental Aircraft Systems (KEAS) Laboratory to characterize quadrotor linear models for wind estimation. The flight experiments were designed to identify models approximating the quadrotor dynamics about the equilibrium flight conditions corresponding to $V_{z_{\rm eq}} = \{0.0, 0.5, 1.0, 1.5, 2.0\}$ m/s. The experiments required exciting the aircraft from each flight equilibrium in calm atmospheric conditions (i.e., $\bm{w}(\bm{x},t) \approx \bm{0}$ m/s) to minimize the impact of exogenous excitations on the system identification process. The input-output measurements used for system identification consisted of pilot-induced, sinusoidal joystick commands and the vehicle's measured dynamic response.  

The system identification experiments were performed in two parts. A first set of experiments were performed to identify the quadrotor's hovering flight dynamics. This required exciting from equilibrium flight the quadrotor's  plunge, yaw, roll and pitch dynamics shown in Figure~\ref{fig:quadrotor_modes}.
A second set of experiments was conducted to identify quadrotor models for constant ascent rates varying between 0.5 and 2 m/s. This involved exciting the quadrotor's roll and pitch dynamics from equilibrium flight conditions corresponding to $V_{z_{\rm eq}}>0$. For the latter case, the plunge and yaw dynamics of the quadrotor were assumed to be well approximated by models identified for hovering flight considering that the vehicle's response to wind perturbations in steady-ascending flight is dominated by roll and pitch motions. Measurements from both sets of system identification experiments were then used to identify the model structures and parameter estimates approximating the quadrotor's dynamics for all five operating conditions specified by $V_{{z}_{\rm eq}}$. 

\subsubsection{Model Structure Determination}
\label{ss:model_structure_determination}

The parameter structure of each model was determined from input-output measurements employing the stepwise regression algorithm described in~\cite{klein2006aircraft}.  Using this approach, a set of postulated regressors, $\bm{\chi} = \{\chi_1,\chi_2,\cdots,\chi_n\}$ is tested one at a time to determine which ones significantly improve the fit of the model
\beq z(k) = a_{0}+\sum_{i = 1}^{m}a_i\chi_{i}(k),\hspace{1cm} k= 1,2,\cdots,N\eeq
 where $z$ is the quadrotor's measured response, $a_0$ is the model bias, $\bm{a} = \{a_0,a_1,\cdots,a_m\}$ is the set of model coefficients associated with $ m $ regressors, and $N$ is the sample size of measurements. How well each model structure fits the observed data as regressors are added or removed is determined using the $F_0$ statistic and coefficient of determination $R^2$ metrics. The $F_0$ statistic quantifies how much each regressor contributes to the fit of the model. The coefficient of determination quantifies how well the model output matches the measured data. Using both metrics, a total of four parameter structures were identified to characterize the quadrotor's plunge, yaw, roll, and pitch dynamics.

\subsubsection{Parameter Estimation}
\label{ss:mm_parameter_estimation}
The model structures determined from step-wise regression were used to initialize the estimation of model parameters using the output error algorithm described in \cite{klein2006aircraft}. The output error algorithm estimates model parameters using the output of the linear aircraft model described by Equation~\reff{eqn:linearized_dynamics} in still air conditions and using the N sample points of measured flight data, which are assumed to be corrupted by sensor noise $\bm{\eta}$. The model and measurements used by the output error method are summarized below:
\beqarr
\frac{d}{dt}\bm{\tilde x} &=& \bm{A}\bm{\tilde{x}}+\bm{B}\bm{\tilde{u}},\hspace{1cm} \tilde{x}(0) = x_0 \label{eqn:oe_state_model}\\ \bm{y} &=& \bm{C}\bm{\tilde{x}}+\bm{D}{\bm{\tilde{u}}} \label{eq:oe_output_model}\\
\bm{z}(k) &=& \bm{y}(k)+\bm{\eta}(k)\hspace{1cm} k = 1,2,\cdots,N  \label{eqn:oe_measurement_model}
\eeqarr
 This formulation of the output error method assumes that the model being identified is free of process noise, making numerical propagation of state measurements possible. Moreover, output error parameter estimation assumes flight measurements to be corrupted with uncorrelated, zero-mean Gaussian noise $\bm{\eta} \in \mathcal{N}(\bm{0},\bm{R}_{\rm Cov})$ such that the covariance matrix of measurement noise  is diagonal, 
 \[\mathrm{Cov}(\bm{\eta}(k))=E[\bm{\eta}(k)\bm{\eta}^{T}(k)]=\bm{R}_{\rm Cov}\]
 Using this framework, parameter estimates are tuned iteratively  while minimizing the cost function,
\beq J = \tfrac{1}{2}\sum^{N}_{i=1}[\bm{y}(k)-\bm{z}(k)]^T\bm{R}_\mathrm{Cov}^{-1}[\bm{y}(k)-\bm{z}(k)]\eeq
 which is the uncertainty-weighted residual between the model output and observation measurements. 
 
 Employing the output error approach, three sets of parameters were estimated and averaged to characterize quadrotor models for hovering and steady vertical ascent conditions. The quadrotor models characterized from averaged parameter estimates were validated using a separate flight test data set collected during system identification experiments. 
\subsubsection{Model Validation}
\label{sss:mm_model_validation}

Linear models approximating steady-flight quadrotor dynamics were validated using input-output data collected separately during system identification flight experiments. The validation process for linear models involved comparing model outputs and state measurements corresponding to pilot-generated excitations using the root-mean-squared error (RMSE) metric: \[RMSE = \sqrt{\frac{1}{N}\sum_{k=1}^{N}(y(k)-z(k))^2}\] where $y$ is the model output, $z$ is the state measurements, and $N$ is the measurement sample size. In general, small RMSE values are indicative of accurate parameter estimates.  Results from the RMSE quantification were used to assess the goodness of each model prior to designing a state observer for wind estimation.

\subsection{Observer Synthesis}
\label{ss:observer_synthesis}

To synthesize observers for wind velocity estimation, the dynamic rigid body wind sensing method presented in \cite{gonzalez2019sensing} was adapted. Therefore, assuming absolute measurements from the GPS antenna and AHRS on board the quadrotor to be available, the output equation, as in \cite{gonzalez2019sensing}, is of the form 
\[
\bm{y} = \mathbb{I}_{12} \tilde{\bm{x}} + \begin{pmat}({.}) \bm{0}_3 \cr \bm{0}_3 \cr\- \mathbb{I}_3  \cr \bm{0}_3 \cr \end{pmat} \bm{V}_{\rm w}
\]
where output measurements of translational velocity are the summation of both air-relative and wind velocity (with identity and zero matrices written in short notation, e.g., $\mathbb{I}_{12} \in \mathbb{R}^{12\times 12}$). The quadrotor's output measurement and identified models were then used to formulate wind-augmented models for the set of operating conditions prescribed by $V_{z_{\rm eq}}$. 

 Wind velocity was estimated using the quadrotor models identified from system identification in a state observer framework. State observers were developed based on wind-augmented models corresponding to each of five equilibrium flight conditions. Each wind-augmented model is obtained by reformulating~\reff{eqn:linearized_dynamics} such that the wind disturbance is part of the wind-augmented state vector: $\bm{x}_{\rm A} = [\bm{\tilde{x}}^T,\bm{w}^T]^T$. Here, as in \cite{gonzalez2019sensing}, variations of wind velocity with respect to time were assumed to vary slowly relative to the dynamics of the quadrotor such that $\frac{d}{dt}\bm{w} \approx \bm{0}$. Therefore,  wind-augmented dynamic models corresponding to each flight equilibrium were defined as follows:
\beq
\frac{d}{dt}\bm{x}_{\rm A} =
\underbrace{\begin{pmat}({|}) \bm{A} & \bm{\Gamma} \cr\-
\bm{0}_{3 \times 12}  & \bm{0}_3 \cr \end{pmat}}_{\bm{A}_{\rm A}} \bm{x}_{\rm A}
+ \underbrace{\begin{pmat}({}) \bm{B} \cr\- \bm{0}_{3 \times 4} \cr \end{pmat}}_{\bm{B}_{\rm A}} \tilde{\bm{u}} \hspace{1cm} 
\bm{y} = \underbrace{\begin{pmat}({...|})
\mathbb{I}_3 & \bm{0}_3 & \bm{0}_3 & \bm{0}_3 & \bm{0}_3 \cr
\bm{0}_3 & \mathbb{I}_3 & \bm{0}_3 & \bm{0}_3 & \bm{0}_3 \cr
\bm{0}_3 & \bm{0}_3 & \mathbb{I}_3 & \bm{0}_3 & \mathbb{I}_3 \cr
\bm{0}_3 & \bm{0}_3 & \bm{0}_3 & \mathbb{I}_3 & \bm{0}_3 \cr
\end{pmat}}_{\bm{C}_{\rm A}} \bm{x}_{\rm A}
\label{eqn:AugmentedDynamics}
\eeq
where $\bm{A}_{\rm A}\in \mathbb{R}^{15\times15}$ is the wind-augmented state matrix, $\bm{B}_{\rm A}\in\mathbb{R}^{15\times4}$ is the wind-augmented input matrix, and $\bm{C}_{\rm A}\in\mathbb{R}^{12\times 15}$ is the wind-augmented output model. 

To verify the observability of the augmented dynamic model, an observability analysis was conducted to determine if wind estimates can be constructed from the model and output measurements. The system is observable if and only if the observability matrix defined below is column-wise full rank.

\[
\bm{\mathcal{O}}(\bm{C}_{\rm A},\bm{A}_{\rm A})
= \begin{pmat}({})
\bm{C}_{\rm A} \cr
\bm{C}_{\rm A} \bm{A}_{\rm A} \cr
\vdots \cr
\end{pmat}
= \begin{pmat}({...|})
\mathbb{I}_3 & \bm{0}_3 & \bm{0}_3 & \bm{0}_3 & \bm{0}_3 \cr
\bm{0}_3 & \mathbb{I}_3 & \bm{0}_3 & \bm{0}_3 & \bm{0}_3 \cr
\bm{0}_3 & \bm{0}_3 & \mathbb{I}_3 & \bm{0}_3 & \mathbb{I}_3 \cr
\bm{0}_3 & \bm{0}_3 & \bm{0}_3 & \mathbb{I}_3 & \bm{0}_3 \cr\-
\bm{0}_3 & \bm{G}_w& \mathbb{I}_3 & \bm{0}_3 & \mathbb{I}_3 \cr
\bm{0}_3 & \bm{0}_3 & \bm{0}_3 & \mathbb{I}_3 & \bm{0}_3 \cr
\bm{0}_3 & \bm{G}_g & - d_w \bm{e}_3 \bm{e}_3^T & \bm{0}_3 & \bm{0}_3 \cr
\bm{0}_3 & \bm{0}_3 & \bm{D}_{m_v} & \bm{D}_{m_\omega} & \bm{0}_3 \cr\-
\vdots & \vdots & \vdots & \vdots & \vdots \cr
\end{pmat}
\]
The analysis shows that the observability matrix is full rank, i.e.,~$\rm{rank}\left[\bm{\mathcal{O}}(\bm{C}_{\rm A},\bm{A}_{\rm A})\right] =15$. Therefore, computing a suitable observer gain matrix $\bm{G}_\mathrm{O}$, state estimates of the following observer will converge to the state of the system~\ref{eqn:AugmentedDynamics}

\beq
\frac{d}{dt}~\widehat{\bm{x}_{\rm A}} =
\bm{A}_{\rm A} \widehat{\bm{x}_{\rm A}}
+ \bm{B}_{\rm A} \bm{u}
+ \bm{G}_\mathrm{O}\left( \bm{y} - \bm{C}_{\rm A} \widehat{\bm{x}_{\rm A}} \right)
\label{eqn:Observer} \eeq
Because the augmented state vector includes the wind velocity, it follows that the state estimator (9) provides a convergent estimate of $\bm{w}$, provided the underlying assumptions hold (e.g., small perturbations from the nominal state).

\section{Experimental Validation of Wind Estimates}
\label{sec:experimental_validation_of_wind_estimates}

\subsection{Field Experiment Setup}
\label{ss:field_experiment_setup}

\begin{figure}[tbh!]
\centering
\subfigure[Gill WindSonic (SA)]{\includegraphics[width = 4.5 cm]{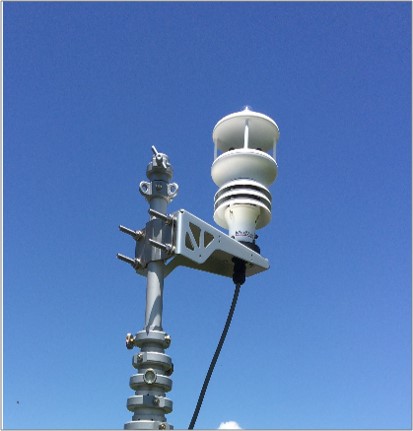}
\label{subfig:Gill_SA_Sensor}}\qquad
\subfigure[ASC LR-SoDAR (SR SoDAR)]{\includegraphics[width = 4.5 cm]{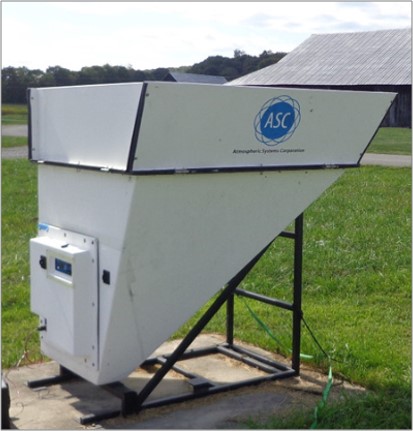}\label{subfig:ASC_SoDAR_Sensor}}\qquad
\subfigure[Remtech PA-0 (SR SoDAR)]{\includegraphics[width = 4.5cm]{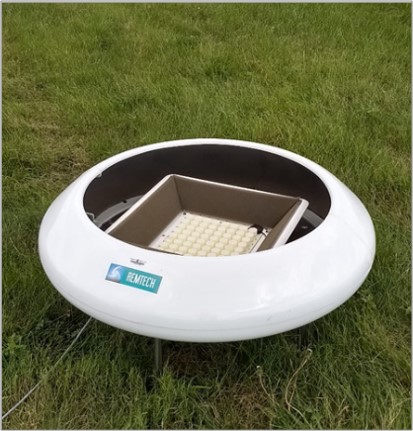}\label{subfig:Remtech_SoDAR_Sensor}} \qquad
\subfigure[Experimental Setup ]{\includegraphics[width = 12 cm]{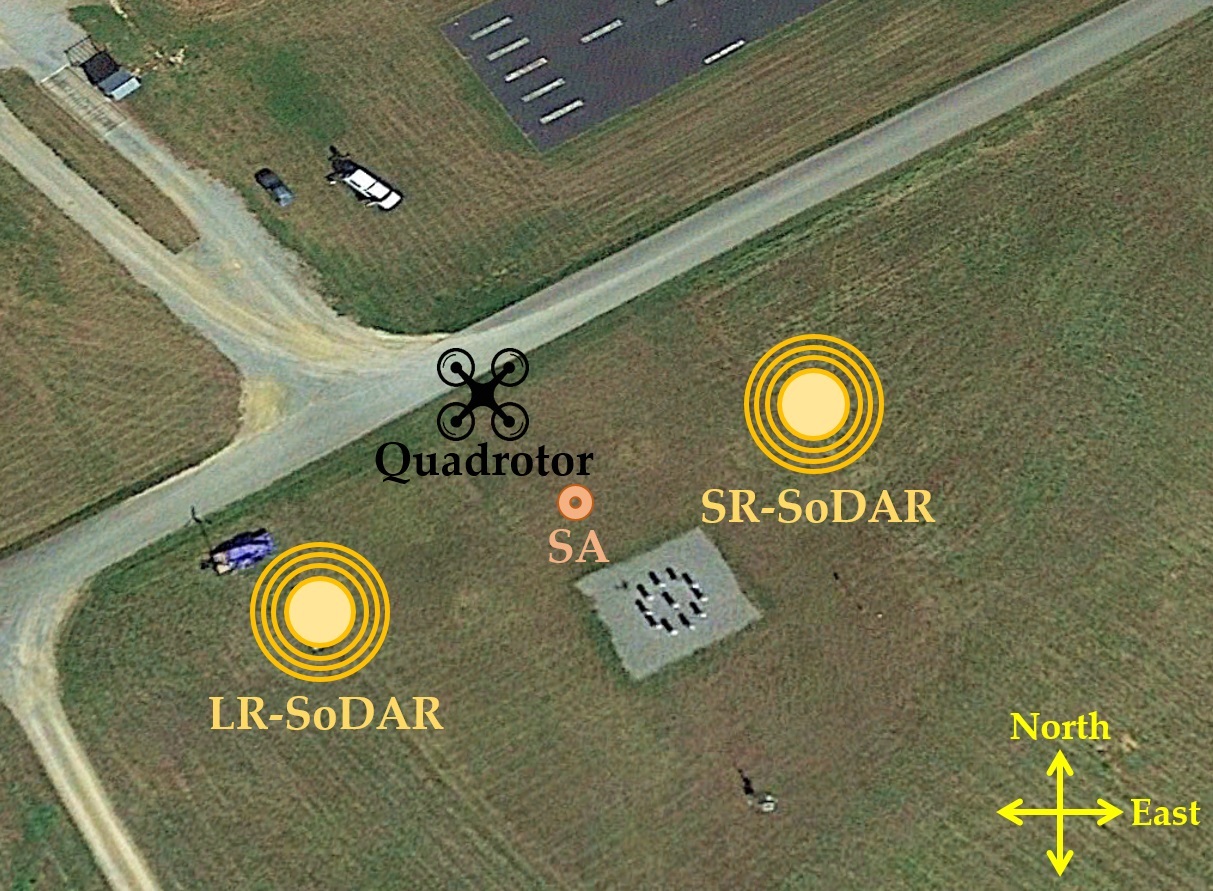}
\label{fig:experiment_setup}}\qquad
\caption{Wind sensors used for comparison of quadrotor wind estimates from 10 to 120 m AGL.}
  	\label{fig:atmospheric_sensors}
  \end{figure}

Field experiments were performed at the KEAS Laboratory on June 5th, 2018  from 9:00 to 20:30 EDT to validate wind estimates from the quadrotor both hovering (i.e., $V_{z_{\rm eq}} = 0$ m/s) and ascending vertically with constant rates,  $V_{z_{\rm eq}} =\{0.5,1.0,1.5,2.0\}$~m/s. During the experiment, data were collected from several in situ and remote sensors shown in Figure~\ref{fig:atmospheric_sensors}. The in situ sensor shown in Figure~\ref{subfig:Gill_SA_Sensor} is a Gill WindSonic (SA) that was mounted on top of a telescoping tower to measure wind velocity at 10~m above ground level (AGL). The remote sensor shown in Figure~\ref{subfig:ASC_SoDAR_Sensor} is an ASC4000i SoDAR (LR SoDAR) and was used to measure  wind velocity from 30 to 400 m AGL.  The remote sensor shown in Figures~\ref{subfig:Remtech_SoDAR_Sensor} is a Remtech SR-SoDAR (SR SoDAR) and measured wind velocity profiles from 10 to 200~m AGL. The performance characteristics of all three sensors are listed in Table~\ref{table:independent_sensors}. The ground setup of all three sensors and the location of flight operations is shown in Figure~\ref{fig:experiment_setup}. Using this sensor setup quadrotor wind velocity estimates were validated.


\begin{table}[tbh!]\footnotesize
\centering
\resizebox{15.5cm}{!}{%
  \begin{tabular}{ccccccccccccccc}
    \toprule
    \multirow{2}{*}{Make/Model} & \multirow{2}{*}{Descriptor}& \multirow{2}{*}{Range} &\multicolumn{2}{c}{Resolution}&&
      \multicolumn{2}{c}{ Accuracy }\\
      \cline{4-5} \cline{7-8}\
       &&& {Spatial} & {Temporal} && {Wind Speed}  & { Wind Direction}    \\
      \bottomrule \midrule
  ASC   &LR-SoDAR &  30 - 410 m &  5 m &30 s & &$\small < 5$ m/s above 2 m/s  &$ 2^\circ$ above 2 m/s   \\
  Remtech PA-0 & SR-SoDAR&  10 - 200 m &  10 m &  300 s&& < 0.2 m/s above 6 m/s  & $3^\circ$ above 2 m/s  \\
 Gill WindSonic &SA&   N/A & N/A &  0.25 s &&  $< 1.0\%$ at 12 m/s&$ 0.5^\circ$ at 12 m/s \\
\midrule
  \bottomrule
  \end{tabular}}
  \caption{Accuracy specifications for sonic anemometer and SoDARs}
  \label{table:independent_sensors}
  \end{table}
\subsection{Ground-Based Observations}
\label{ss:Ground-Based Observations}

 To validate quadrotor wind estimates with ground-based sensors, sonic anemometer and SoDAR observations were first compared starting from mid morning until evening on June 5th, 2018. Wind measurements from the SA and SR-SoDAR were sampled at 10 m AGL from 13:00 to 20:30 EDT. Measurements from the LR-SoDAR  were not included in comparisons at 10 m AGL since it only measures above 30 m AGL. Wind observations from the SR-SoDAR and LR-SoDAR were also compared at 30, 70, and 120 m AGL from 9:30 to 20:30 EDT. In this process, sonic anemometer and LR-SoDAR measurements were averaged over 300 second intervals for uniform comparison with SR-SoDAR measurements. The agreement across sensors was quantified using as metrics the mean bias error (MBE) and RMSE of wind observations while assuming wind spatial variations to be negligible over the lateral footprint of the sensors; see Figure~\ref{fig:atmospheric_sensors}. Results from the comparison were used to aid the assessment of quadrotor wind estimates.  

\subsection{Wind Data Measurements}
\label{ss:wind_data_measurements}

Following the comparison among ground-based sensors, wind observations from the sonic anemometer and SoDAR sensors were used to validate quadrotor wind estimates from hovering and steady-ascending flight operations. Wind estimates from the quadrotor hovering at 10 m AGL were validated using SA and SR-SoDAR wind observations. Quadrotor wind profiles extending from 10 to 120 m AGL were validated using wind measurements from the SR- and LR-SoDARs. Considering that the quadrotor samples the atmosphere in continuous vertical ascent, quadrotor wind velocity estimates were averaged over 10-m intervals for comparison with SoDAR measurements. Outcomes from both comparisons were used to benchmark the performance of quadrotor wind estimation relative to conventional atmospheric sensors.

\section{Results}
\label{sec:results}

\subsection{System Identification}

\subsubsection{Model Structure Determination}
\label{ss:model_structure_determination_results}
 The quadrotor flight dynamic model for hovering and steady ascending flight is decomposed into four sub-models that describe plunging, yawing, rolling, and pitching motion. Table~\ref{table:model_structure_results} shows all four model forms and associated parameters. The plunge model is a system of two first-order ordinary differential equations parameterized by propulsive and damping parameters. The yaw model is a system of two first-order ordinary differential equations with rotational damping and torque parameters. Finally, the roll and pitch models are systems of four first-order ordinary differential equations.


\begin{table}[tbh!]
\centering
  \begin{tabular}{ccl}
    \toprule 
    Model && \multicolumn{1}{c}{Parameter Structure}  \\
      \toprule \midrule
Plunge && $ \left(\overliner{c} \dot{z}\\ \dot{w}\earr\right) = \left(\overliner{cc} 0 & 1\\ 0 & Z_{w} \earr\right)\left(\overliner{c} z \\ w\earr\right)+\left(\overliner{c} 0 \\ Z_{\delta}\earr\right)\delta_{\rm plunge}$\\ \\
Yaw & & $\left(\overliner{c} \dot{\psi}\\ \dot{r}\earr\right) = \left(\overliner{cc}0 & 1\\ N_\psi & N_{r} \earr\right)\left(\overliner{c}\psi \\ r\earr\right)+\left(\overliner{c} 0 \\ N_{\delta}\earr\right)\delta_{\rm yaw}$\\ \\
Roll & &$\left(\overliner{c}\dot{y}\\\dot{\phi} \\\dot{v}\\\dot{p} \earr\right) = \left(\overliner{cccc} 0& 0&1& 0\\0 &0 &0 &1 \\ 0&Y_{\phi} &Y_{v}&0\\0& L_{\phi}& 0& L_{p}\earr\right) \left(\overliner{c}y\\\phi \\v\\p \earr\right)+\left(\overliner{c}0\\ 0\\0\\L_{\delta} \earr\right)\delta_{\rm roll}$ \\ \\
Pitch & & $\left(\overliner{c}\dot{x}\\ \dot{\theta} \\ \dot{u} \\ \dot{q} \earr\right) = \left(\overliner{cccc} 0&0 &1 & 0\\ 0 & 0 & 0 & 1\\0 &X_{\theta} &X_{u}&0\\ 0&M_{\theta}&0&M_{q}\earr\right) \left(\overliner{c}x\\ \theta \\ u \\ q \earr\right)+\left(\overliner{c}0\\ 0 \\ 0 \\ M_{\delta}  \earr\right)\delta_{\rm pitch}$\\

\midrule \bottomrule
  \end{tabular}
  \caption{The plunge, yaw, roll and pitch model structures of the quadrotor determined from system identification flight experiments and step-wise regression algorithm presented in~\cite{klein2006aircraft}.}
  \label{table:model_structure_results}
  \end{table}


\subsubsection{Parameter Estimation}
\label{ss:parameter_estimation}
Three sets of quadrotor model parameters were estimated for each of five equilibrium flight conditions. Each set of parameters was estimated using the model structures determined from step-wise regression and the output error algorithm described in Section~\ref{ss:aircraft_system_identification}. Model parameters for the plunge, yaw, roll, and pitch model structures were first estimated for hovering flight conditions (i.e., $\bm{v} = \bm{0}$ and $\bm{\omega}=\bm{0}$). Subsequently, roll and pitch model parameters were estimated for constant vertical ascent flight conditions varying from 0.5 to 2.0~m/s. We assume the plunge and yaw model parameters to be invariant with vertical ascent rate. Model parameter estimates for each flight equilibrium were averaged to obtain nominal models for wind estimation. Averaged parameter estimates and standard error (SE) values for plunge and yaw models are listed in Table~\ref{table:plunge_yaw_parameters}. Additionally, averaged roll and pitch  model parameters and standard error values are listed for hovering and ascending flight conditions in Tables~\ref{table:oe_roll_parameters} and \ref{table:oe_pitch_parameters}, respectively.  
\begin{table}[tbh!]
\centering
  \begin{tabular}{cccccccccccccc}
    \toprule
     \multirow{2}{*}{Speed} & \multicolumn{4}{c}{Plunge Model }&&
      \multicolumn{4}{c}{ Yaw Model } \\
      \cline{2-5} \cline{7-9}   \\[-1em]
     &Parameter & Value & SE & Units &   & Parameter  & Value & SE & Units \\
      \toprule \midrule
  \multirow{3}{*}{0-2 ~m/s}&$Z_w$    & -0.55  &0.28 &1/s &&$N_{\psi}$& -1.71& 0.41 &1/s$^2$\\&$Z_\delta$  & -1.71  &0.79 & 1/kg & &$N_r$&-0.84   & 0.53&  1/s  \\

  &--  & -- &-- & --&& $N_\delta$ &2.41   & 1.18  &$1/(\rm kg\cdot m^2)$ \\
        \midrule
    \bottomrule
  \end{tabular}
  \caption{Nominal plunge and yaw model parameter estimates.}
    \label{table:plunge_yaw_parameters}
  \end{table}

\begin{table}[tbh!]
\centering
\resizebox{15.5cm}{!}{\begin{tabular}{cccccccccccccccccccccccccc}
      \toprule 
     Pitch Model &\multicolumn{2}{c}{ 0.0 m/s }&& \multicolumn{2}{c}{ 0.5 m/s }&& \multicolumn{2}{c}{ 1.0 m/s }&& \multicolumn{2}{c}{ 1.5 m/s }&& \multicolumn{2}{c}{ 2.0 m/s }&&\multirow{2}{*}{Units}\\
      \cline{2-3} \cline{5-6} \cline{8-9} \cline{11-12} \cline{14-15} \\[-1em]
     Parameters&Value& SE & &Value&  SE& &  Value & SE && Value& SE&& Value& SE\\
      \toprule\midrule
  $Y_\phi$ & 3.28  &0.37 && 2.91   & 0.34&& 4.73   & 0.87 && 4.68  & 0.21&&6.62&0.63&& m/s$^2$  \\
  $Y_v$    & -0.49 &0.68 && -0.31  & 0.04&& -0.70  & 2.33 && -0.62 & 0.14&&-1.06&0.25&& 1/s  \\
  $L_\phi$ & -4.54 &4.17 && -3.95  & 0.12&& -5.87  & 2.55 && -4.07 & 0.26&&-5.92&0.10&& 1/s$^2$\\
  $L_p$    & -1.09 &2.62 && -1.15  & 0.22&& -1.62  & 1.99 && -0.82 & 0.23&&-1.80&1.17&& 1/s\\
  $L_\delta$  & 4.62  &3.55 && 5.76   & 0.32&& 8.52   & 2.28 && 6.27  & 0.31&&9.68&0.65&&  $1/(\rm kg\cdot m^2)$ \\
  \midrule
\bottomrule
\end{tabular}}
\caption{Nominal roll model parameter estimates.}
\label{table:oe_roll_parameters}
\end{table}

\begin{figure}[tbh!]
    \centering
    \includegraphics[width =\textwidth]{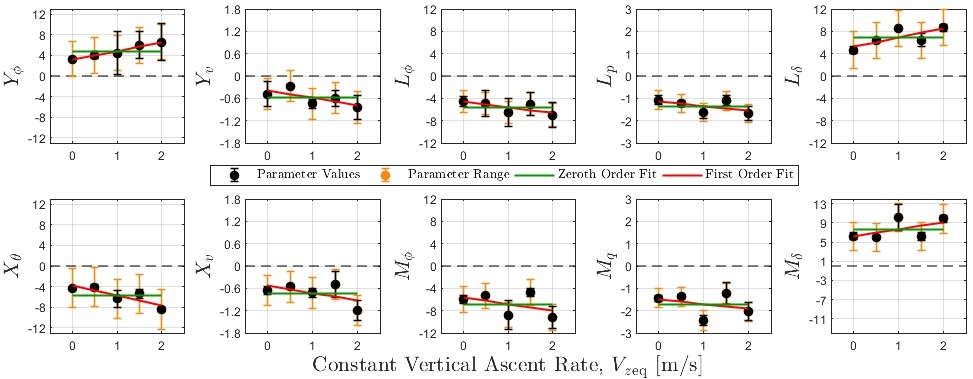}
    \caption{Quadrotor roll and pitch model parameter estimates.  }
    \label{fig:quadrotor_roll_pitch_parameters}
\end{figure}

  \begin{table}[tbh!]
\centering
  \resizebox{15.5cm}{!}{\begin{tabular}{cccccccccccccccccccccccccc}
      \toprule 
     Pitch Model &\multicolumn{2}{c}{ 0.0 m/s }&& \multicolumn{2}{c}{ 0.5 m/s }&& \multicolumn{2}{c}{ 1.0 m/s }&& \multicolumn{2}{c}{ 1.5 m/s }&& \multicolumn{2}{c}{ 2.0 m/s }&\multirow{2}{*}{Units}\\
      \cline{2-3} \cline{5-6} \cline{8-9} \cline{11-12} \cline{14-15} \\[-1em]
     Parameters&Value& SE & &Value&  SE& &  Value & SE && Value& SE&& Value& SE\\
      \toprule\midrule
 $X_\theta$ & -4.03  &0.10 &&-3.94  &0.12 && -6.27   &0.78  &&-5.48  &0.14 && -8.02&0.68& m/$s^2$  \\
 $X_u$    & -0.71    &0.56 &&-0.61  &0.08 && -0.80   &0.19 && -0.67 &0.08 && -1.24&0.28& 1/s  \\
 $M_\theta$ & -6.23  &1.67 &&-5.20  &0.11 && -8.63   &2.64 && -4.44 &0.23 && -7.78&2.69& 1/s$^2$\\
 $M_q$    & -1.46    &0.87 &&-1.42  &0.35 && -2.63   &0.65 && -1.27 &0.50 && -2.09&0.84& 1/s\\
 $M_\delta$  & 6.61     &0.36 &&6.32   &0.28 && 10.80   &1.98 && 6.81  &0.40 && 10.70&0.64&  $1/(\rm kg\cdot m^2)$ \\
  \midrule
    \bottomrule
  \end{tabular}}
  \caption{Nominal pitch model parameter estimates.}
    \label{table:oe_pitch_parameters}
  \end{table}
The dependence on vertical ascent rate was also characterized for roll and pitch quadrotor parameters. Results from this characterization are shown in Figure~\ref{fig:quadrotor_roll_pitch_parameters} where roll and pitch model parameters are plotted as a function of ascent rate. Each parameter estimate appears with absolute error bars, colored in black, representing the range of estimates obtained from the three experimental data sets. Orange-colored bars were also included to denote minimum and maximum values across all five ascent rates.  Zeroth- and first-order polynomials were fit to the parameter estimates as a function of ascent rate. The first-order fit, on the other hand, characterizes the trend in parameter values with respect to ascent rate. Note that only a subset of parameters exhibit clear trends with respect to ascent rate. It is possible that these local, small-perturbation models do exhibit high sensitivity to ascent rate, as suggested by Figure~\ref{fig:quadrotor_roll_pitch_parameters}. If so, then these results may suggest flight regimes to be avoided when estimating wind velocity from platform motion; regions of high parameter sensitivity may produce less accurate wind estimates.

For the aircraft and dynamic model considered here, the parameters vary less at lower ascent rates (0.5 m/s or less). Thus, one might expect more accurate wind measurements during slower climbs. It is possible, however, that the variation in parameter estimates is an artifact of the data collection method for system identification. At higher climb rates, it is more difficult to manually generate the rich and precisely timed excitation signals needed for model identification. An automated approach to system identification may improve the repeatability of parameter estimates.

\subsubsection{Model Validation}
\label{ss:model_validation}
\begin{figure}[tbh!]
\centering 

\subfigure[Plunge model validation]{\includegraphics[width = 77mm]{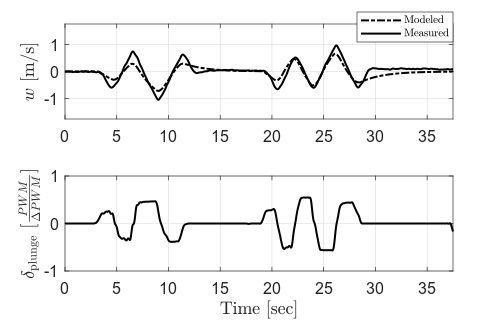}
\label{subfig:plunge_model0ms}}
\subfigure[Yaw model validation]{\includegraphics[width = 77mm]{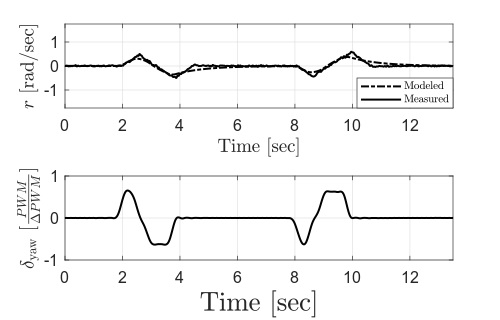}\label{subfig:yaw_model0ms}}\\
\subfigure[Roll model validation]{\includegraphics[width = 77mm]{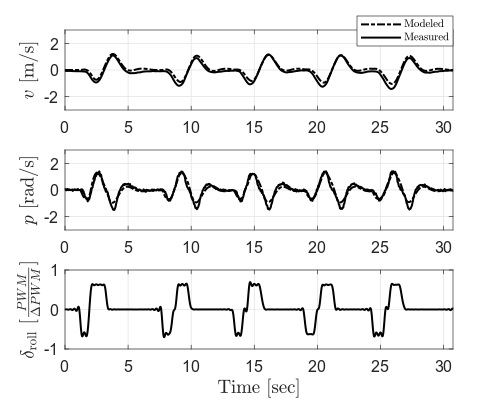}
\label{subfig:roll_model0ms}}
\subfigure[Pitch model validation]{\includegraphics[width = 77mm]{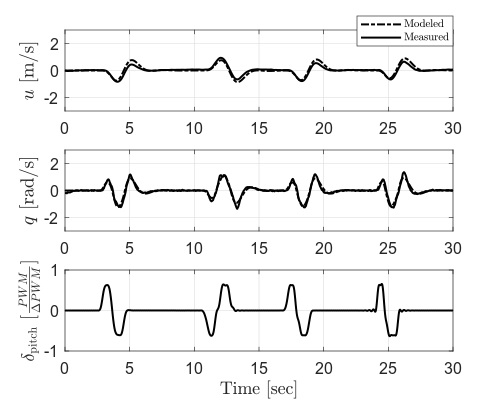}\label{subfig:pitch_model0ms}}
\caption{ Validation of plunge, yaw, roll, and pitch models identified for quadrotor hovering flight.}
  	\label{fig:SID_validation}
  \end{figure}

Models characterized from step-wise regression and output error methods were validated by comparing the model output and aircraft's response to an excitation input. Agreement between the model output and measured response is compared  using the RMSE metric discussed in Section~\ref{sss:mm_model_validation}. Results from this validation are shown in Figure~\ref{fig:SID_validation} for the plunge, yaw, roll, and pitch models characterized for hovering flight $V_{z_{\rm eq}} = 0$. Results from the RMSE assessment for the plunge and yaw models are shown in Table~\ref{table:roll_pitch_SID_validation_Results}. The RMSE results for the pitch and roll models associated constant vertical ascent rates ranging between 0 and 2 m/s are also shown in Table~\ref{table:roll_pitch_SID_validation_Results}.

\begin{table}[tbh!]
\centering
\resizebox{15.5cm}{!}{%
  \begin{tabular}{cccccccccccccccccc}
    \toprule
     \multirow{2}{*}{Ascent} &\multicolumn{3}{c}{ Plunge Model }&& \multicolumn{3}{c}{ Yaw Model }&& \multicolumn{3}{c}{ Roll Model }&&
      \multicolumn{3}{c}{ Pitch Model} &\\
      \cline{2-4} \cline{6-8} \cline{10-12} \cline{14-16}   \\[-1em] Rate &Par. & RMSE & Units   &   & Par. & RMSE & Units &   & Par. & RMSE & Units &   & Par. & RMSE & Units  \\
      \toprule\midrule
  \multirow{2}{*}{0 m/s}  &\multirow{2}{*}{$w$}&\multirow{2}{*}{0.44}&\multirow{2}{*}{m/s}&& \multirow{2}{*}{$r$}&\multirow{2}{*}{2.59}&\multirow{2}{*}{rad/s}& &$v$ & 0.23   & m/s& &$u$   & 0.12  & m/s \\
  &&&&&&&& &$p$ &0.39   & rad/s& &$q$   & 0.19  & rad/s \\ \midrule
    \multirow{2}{*}{0.5 m/s}&\multirow{2}{*}{$w$}&\multirow{2}{*}{0.44}&\multirow{2}{*}{m/s}&& \multirow{2}{*}{$r$}&\multirow{2}{*}{2.59}&\multirow{2}{*}{rad/s}& &$v$ & 0.31   & m/s& &$u$   & 0.59  & m/s \\
  & &&&&&&&& $p$ & 0.21   & rad/s& &$q$   & 0.31  & rad/s \\ \midrule
  \multirow{2}{*}{1.0 m/s} &\multirow{2}{*}{$w$}&\multirow{2}{*}{0.44}&\multirow{2}{*}{m/s}&& \multirow{2}{*}{$r$}&\multirow{2}{*}{2.59}&\multirow{2}{*}{rad/s}& &$v$ & 0.73   & m/s& &$u$   & 0.38  & m/s \\
  &&&&&&&& &$p$ & 0.90  & rad/s& &$q$   &  0.37 & rad/s \\ \midrule
    \multirow{2}{*}{1.5 m/s}&\multirow{2}{*}{$w$}&\multirow{2}{*}{0.44}&\multirow{2}{*}{m/s}&& \multirow{2}{*}{$r$}&\multirow{2}{*}{2.59}&\multirow{2}{*}{rad/s}& &$v$ & 0.38   & m/s& &$u$   &  0.46 & m/s \\
  & &&&&&&&&$p$ &0.48   & rad/s& &$q$   & 0.51 & rad/s \\\midrule
    \multirow{2}{*}{2.0 m/s} &\multirow{2}{*}{$w$}&\multirow{2}{*}{0.44}&\multirow{2}{*}{m/s}&& \multirow{2}{*}{$r$}&\multirow{2}{*}{2.59}&\multirow{2}{*}{rad/s}& &$v$ &0.48    & m/s& &$u$   &  0.37 & m/s \\
  &&&&&&&& &$p$ & 0.71  & rad/s& &$q$   & 0.28  & rad/s \\
        \midrule
    \bottomrule
  \end{tabular}}
  \caption{Validation results for plunge, yaw, roll and pitch models.}
    \label{table:roll_pitch_SID_validation_Results}
  \end{table}

\subsection{Comparison of Wind Velocity Measurements}

\subsubsection{Sonic Anemometer and SoDAR Comparison}
\label{s:SoDAR_and_Sonic_Anemometer_Comparison}

\begin{figure}[tbh!]
\centering
\subfigure[]{\includegraphics[width =\textwidth]{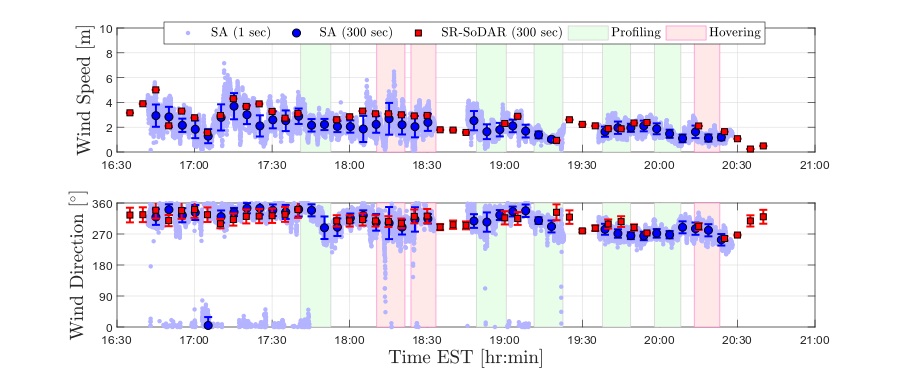}\label{subfig:wind_measurements_10m}}
\subfigure[]{\includegraphics[width =\textwidth]{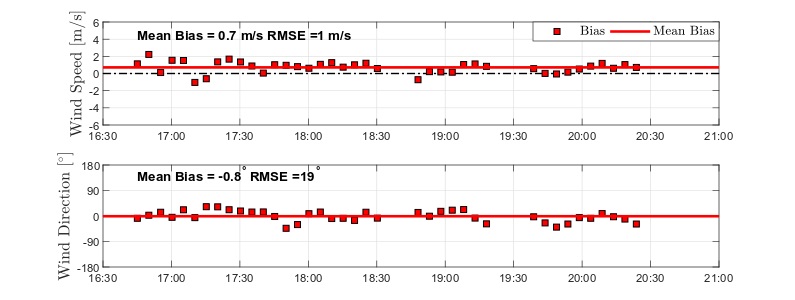}\label{subfig:wind_bias_10m}}
\caption{ Comparison of wind speed and wind direction observations collected from the 
 sonic anemometer and SR-SoDAR at 10 m AGL on June, 5th 2018 from 16:30 to 20:30 EDT. }
  	\label{fig:comparison_bias_10m}
  \end{figure}

Sonic anemometer and SR-SoDAR wind observations were collected from 15:00 to 20:30~EDT to assess their agreement at 10 m AGL. Observations were compared using the mean bias error and RMSE metrics described in Section~\ref{ss:Ground-Based Observations} using sonic anemometer observations as reference.  During the sampling period, prevailing wind conditions were observed to be from the northwest direction with wind speeds ranging from 0 and 6 m/s as shown in Figure~\ref{subfig:wind_measurements_10m}. Results from the comparison are tallied in Table~\ref{table:wind_velocity_observations_10m}. The mean bias error of SR-SoDAR wind speed and wind direction observations was measured to be 0.7 m/s and -0.8$^\circ$, respectively (see Figure~\ref{fig:comparison_bias_10m}). The corresponding RMSE of wind speed and wind direction measurements was determined to be 1.0 m/s and 19.0$^\circ$, respectively. These results were used to assess the accuracy wind estimates from the quadrotor hovering at 10 m AGL as well.

\begin{table}[tbh!]
\centering

  \begin{tabular}{cccccccccccccc}
    \toprule
     \multirow{2}{*}{Sensor} & \multirow{2}{*}{Height}& \multicolumn{3}{c}{Wind Speed}&\multirow{2}{*}{Units}&
      \multicolumn{3}{c}{ Wind Direction} &\multirow{2}{*}{Units}\\
      \cline{3-5} \cline{7-9}   \\[-1em]  & &Mean & MBE& RMSE& & Mean &  MBE & RMSE     \\
      \toprule\midrule
  SA &\multirow{2}{*}{10 m} &2.0 &\multirow{2}{*}{0.7}&\multirow{2}{*}{1.0} & m/s&  284.1 & \multirow{2}{*}{-0.8}&\multirow{2}{*}{19.0} & $^\circ$  \\
  SR-SoDAR & &2.7 & & & m/s & 320.0& & & $^\circ$ \\
    \midrule
    \bottomrule
  \end{tabular}
 \caption{Comparison of wind speed and wind direction observations collected from the 
 sonic anemometer and SR-SoDAR at 10 m AGL on June, 5th 2018 from 16:30 to 20:30 EDT.}
\label{table:wind_velocity_observations_10m}
\end{table}

\subsubsection{SoDAR Comparison}
\label{SoDAR Comparison}

\begin{figure}[tbh!]
\centering
\subfigure[]{\includegraphics[width =13.5cm]{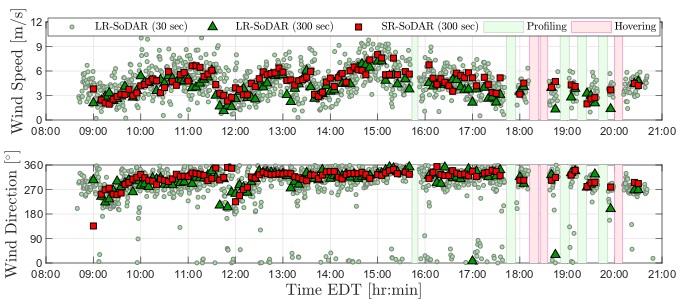}\label{subfig:ASCREM110m}}
\subfigure[]{\includegraphics[width =13.5cm]{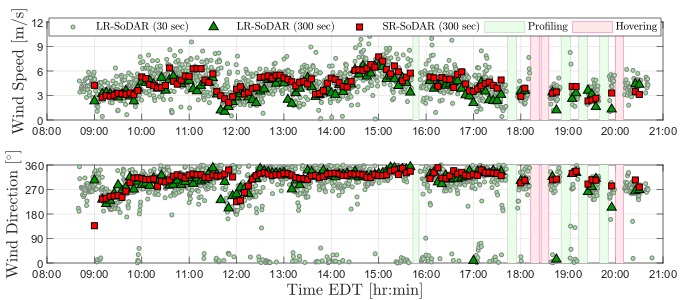}\label{subfig:ASCREM70m}}
\subfigure[]{\includegraphics[width =13.5cm]{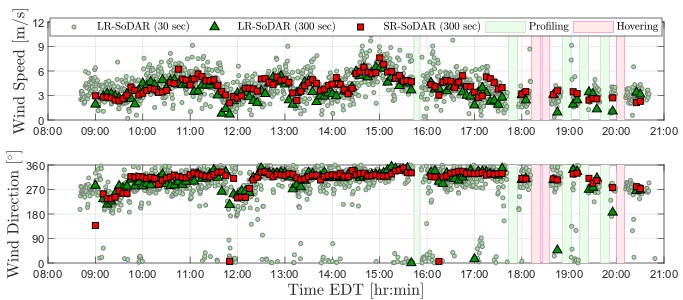}\label{subfig:ASCREM30m}}
\caption{ Wind speed and wind direction observations collected from SoDARs at (a) 110~m AGL , (b) 70~m AGL, and (c) 30~m AGL on June 5th, 2018 from 9:00 to 20:30 EDT. }
  	\label{fig:SoDAR_Measurements}
  \end{figure}

\begin{figure}[tbh!]
\centering
\subfigure[]{\includegraphics[width =13.5cm]{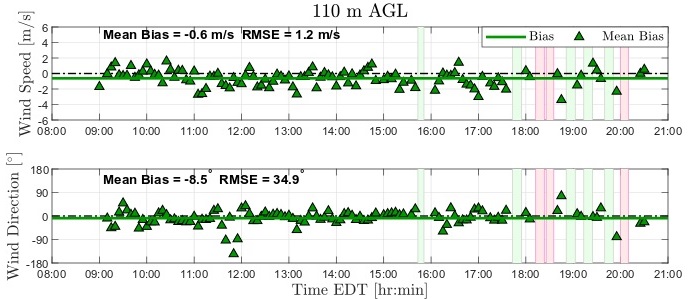}\label{subfig:ASCREM110m_bias}}
\subfigure[]{\includegraphics[width =13.5cm]{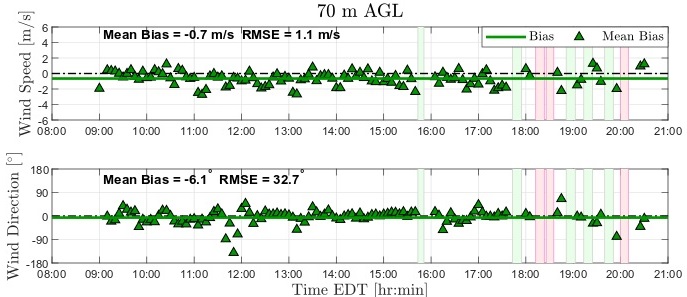}\label{subfig:ASCREM70m_bias}}
\subfigure[]{\includegraphics[width = 13.5cm]{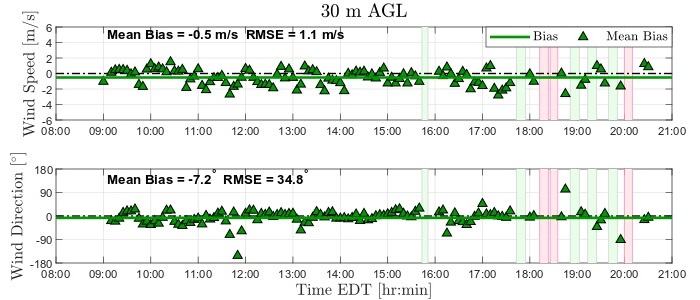}\label{subfig:ASCREM30m_bias}}
\caption{ Comparison of wind speed and wind direction observations collected from the SR- and LR-SoDAR at (a) 110~m AGL, (b) 70~m AGL, and (c) 30~m AGL on June 5th, 2018 from 9:00 to 20:30 EDT.}

  \label{fig:SoDAR_bias_measurements}
  \end{figure}

Measurements of wind speed and wind direction collected from the LR- and SR-SoDARs were compared from 9:00 to 20:30 EDT to assess agreement across wind observations at 30, 70, and 110~m~AGL. Wind conditions during the sampling period were prevalent from the northwest direction with five-minute averages of wind speeds ranging from 1.0 to 8.0~m/s and increasing with height (see Figure~\ref{fig:SoDAR_Measurements}). Results from the comparison of SoDAR wind observations are shown both in Figure~\ref{fig:SoDAR_bias_measurements} and Table~\ref{table:wind_velocity_observations_30_110m} using observations from the SR-SoDAR as reference. The maximum mean bias errors for wind speed and wind direction were observed to be -0.7~m/s and $-7.2^\circ$ at 70 and 30~m AGL, respectively. The maximum RMSE values, on the other hand, were observed for to be 1.2~m/s and $34.9^\circ$ at 110~m AGL. Note that the RMSE values of wind speed and direction agree to within 10\% at all three altitudes. Combined, these results were used to assess the accuracy of quadrotor wind profile estimates.

\begin{table}[tbh!]
\centering

  \begin{tabular}{cccccccccccccc}
    \toprule
     \multirow{2}{*}{Sensor} & \multirow{2}{*}{Height}& \multicolumn{3}{c}{Wind Speed}&\multirow{2}{*}{Units}&
      \multicolumn{3}{c}{ Wind Direction} &\multirow{2}{*}{Units}\\
      \cline{3-5} \cline{7-9}   \\[-1em]  & &Mean & MBE & RMSE& & Mean & MBE & RMSE \\
      \toprule\midrule
  SR-SoDAR &\multirow{2}{*}{30 m} &4.1 &\multirow{2}{*}{-0.5}&\multirow{2}{*}{1.1} & m/s&  305.6 & \multirow{2}{*}{-7.2 }&\multirow{2}{*}{34.8} & $^\circ$  \\
  LR-SoDAR & &3.6 && & m/s & 298.5& & & $^\circ$ \\
    \midrule
  SR-SoDAR &\multirow{2}{*}{70 m} &4.5 & \multirow{2}{*}{-0.7}&\multirow{2}{*}{1.1} & m/s&  310.3 &  \multirow{2}{*}{-6.1}&\multirow{2}{*}{32.7} & $^\circ$  \\
  LR-SoDAR & &3.8 & & & m/s & 301.3& & & $^\circ$ \\
    \midrule
      SR-SoDAR &\multirow{2}{*}{110 m} &4.6&\multirow{2}{*}{-0.6}&\multirow{2}{*}{1.2} & m/s&  312.5&  \multirow{2}{*}{-0.9}&\multirow{2}{*}{34.9} & $^\circ$  \\
  LR-SoDAR & &4.0 & & & m/s & 300.6 & & & $^\circ$ \\
    \midrule
    \bottomrule
  \end{tabular}
 \caption{Comparison of wind speed and wind direction observation from the 
 sonic anemometer and SR-SoDAR at 10 m AGL collected on June, 5th 2018 from 16:30 to 20:30 EDT.}
\label{table:wind_velocity_observations_30_110m}
\end{table}

\subsubsection{Validation of Quadrotor Wind Estimates}

\begin{figure}[tbh!]
\centering

\subfigure[]{\includegraphics[width = 120mm]{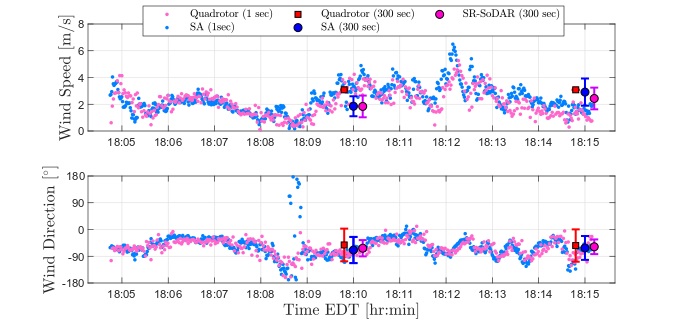}
\label{subfig:Hove1}}
\subfigure[]{\includegraphics[width = 120mm]{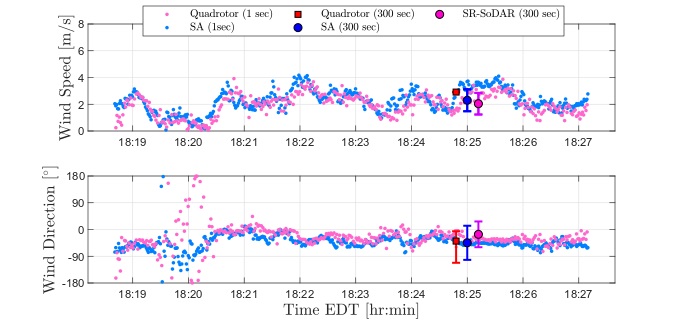}
\label{subfig:Hover1818}}
\subfigure[]{\includegraphics[width = 120mm]{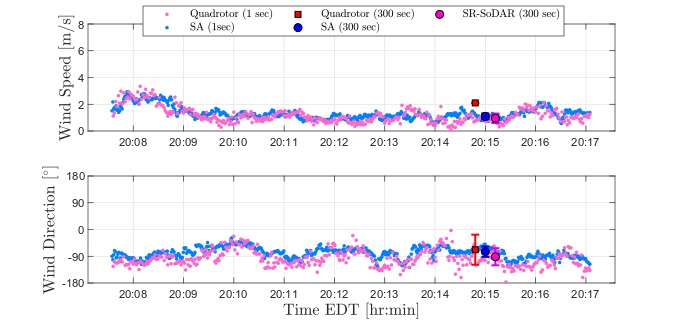}
\label{subfig:HoverC}}
\caption{Wind speed and direction from the SA, SR-SoDAR, and quadrotor at 10~m~AGL.  For clarity of comparison results, quadrotor and SoDAR five-minute averages were offset by 12 seconds.}

  	\label{fig:hover_measurements}
  \end{figure}

Quadrotor wind estimates at 10 m AGL were validated using sonic anemometer and SR-SoDAR wind observations. The validation of quadrotor wind estimates was conducted for three flights occurring between 18:05 and 20:17 EDT. As shown in Figure~~\ref{subfig:wind_measurements_10m} with rose-colored vertical bands, the time lapse of each flight was approximately 10 minutes. To validate quadrotor wind estimates, the bias error of five-minute averages was determined using sonic anemometer and SR-SoDAR observations as reference. Biases of five-minute estimates were then averaged to determine the nominal accuracy of quadrotor wind estimates. Results from this analysis are shown in Figure~\ref{fig:hover_measurements} and Table~\ref{table:SA_SRSoDAR_bias}. For quadrotor wind speed estimates, the absolute value average of mean bias errors and RMSE values were measured to be 0.3 m/s and 1.0 m/s relative to sonic anemometer and SR-SoDAR, respectively. Wind direction quadrotor estimates had absolute value averages of absolute errors and RMSE values equal to $9.9^\circ$ relative to sonic anemometer and SR-SoDAR measurements as well. Therefore, quadrotor wind estimates from hovering flight were characterized by an accuracy comparable to that of conventional ground-based wind sensors.   

\begin{table}[tbh!]
\centering

  \begin{tabular}{cccccccccccccc}
    \toprule
     \multirow{2}{*}{Flight Time} & \multirow{2}{*}{Height}& \multicolumn{2}{c}{Wind Speed Bias Error}&&
      \multicolumn{2}{c}{Wind Direction Bias Error } \\
      \cline{3-4} \cline{6-7}   \\[-1em]  &  &SA & SR-SoDAR& &  SA& SR-SoDAR&     \\
      \toprule\midrule
  \multirow{2}{*}{18:05-18:15 EDT} &\multirow{2}{*}{10 m} & 0.0 m/s & 1.2 m/s && $7.0^\circ$  &$-4.0^\circ$  \\ & & 0.5 m/s&0.7 m/s && $4.0^\circ$&$-6.0^\circ$ 
  \\ \midrule
   18:19-18:27 EDT & 10 m  &-0.5 m/s&-0.8 m/s&&$4.7^\circ$&$12.7^\circ$  \\ \midrule 
   20:08-20:17 EDT&10 m& - 0.1 m/s&-1.2 m/s&& $-23.6^\circ$ &$-17.0^\circ$
   \\
        \midrule
       \multicolumn{2}{c}{\textbf{Absolute Value Average}} & 0.3 m/s &  1.0 m/s && $9.9^\circ$ & $9.9^\circ$& \\\midrule
    \bottomrule
  \end{tabular}
 \caption{Comparison of five-minute averages of wind speed and wind direction observation from the 
 quadrotor, sonic anemometer, and SR-SoDAR at 10 m AGL collected on June, 5th 2018 from 18:05 to 20:17 EDT.}
    \label{table:SA_SRSoDAR_bias}
  \end{table}

 To validate quadrotor wind estimates while ascending vertically at constant rates varying between 0.5 and 2 m/s SoDAR wind observations were used. Preliminary results from validation experiments revealed distinct anomalies in SoDAR wind measurements during quadrotor operations, which took place from 15:30 to 20:00~EDT as shown in Figure~\ref{fig:SoDAR_Measurements} with green-colored vertical bands. During quadrotor operations, wind observations were persistently dropped or not recorded by the SR-SoDAR. Wind observations from the LR-SoDAR, on the other hand, reported an abrupt change in wind speed and wind direction. These anomalies prompted additional investigation to determine the impact of quadrotor operations on the reliability of SoDAR wind observations.

To determine the impact of quadrotor operations on the reliability of SoDAR observations, the two-part assessment detailed in Appendix~\ref{s:a_reliability_study_of_SoDAR_Wind_Measurements} was conducted across the duration of validation experiments. Examination of the spatial footprint of quadrotor operations showed the quadrotor entering the sampling volume of SoDAR observation at approximately 60 m AGL. Both the description of the analysis and results are found in Appendix~\ref{s:a_reliability_study_of_SoDAR_Wind_Measurements}. The quadrotor flying through the sampling volume of SoDAR observations correlated with both an increase noise intensity and a decrease in signal-to-noise ratio corresponding to wind measurements. As shown in Figures~\ref{fig:LR-SoDAR_SR} and \ref{fig:SR_Noise}, the noise intensity of wind velocity measurements increased for $\bm{u}$ and $\bm{v}$ components during quadrotor operations. Correspondingly, the signal-to-noise ratio dropped during quadrotor operations as shown in Figures~\ref{fig:LR-SoDAR_SNR}-\ref{fig:SR-SoDAR_SNR}. Based on these observations, making time-synchronized comparisons of quadrotor and SoDAR wind measurements at the same location for validation purposes is infeasible. 

To circumvent corrupted SoDAR wind observations, quadrotor wind profiles were validated using SoDAR measurements collected 15 minutes prior to and following quadrotor operations. As a result, fair comparisons of quadrotor and SoDAR wind measurements could only be made during period of low wind variability. Otherwise, quadrotor wind estimates were ruled inconclusive. Wind variability was considered to be low when SoDAR wind observations were in general agreement for the duration of the comparison period. We also note that the assessment of quadrotor estimates was performed qualitatively instead of quantitatively due to the non-uniformity of quadrotor SoDAR wind observations across time and space. Consequently, only a subset of wind estimates was validated successfully applying this criteria.  

\begin{figure}[tbh!]
\centering
\subfigure[]{\includegraphics[width = 67mm]{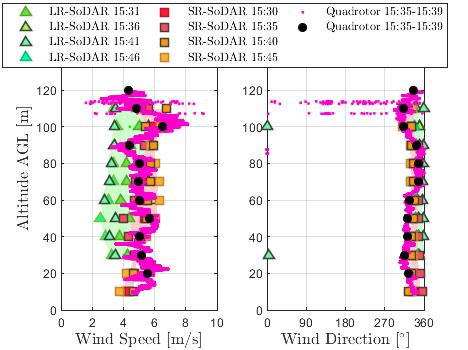}
\label{subfig:0p5ms1}} 
\subfigure[]{\includegraphics[width = 67mm]{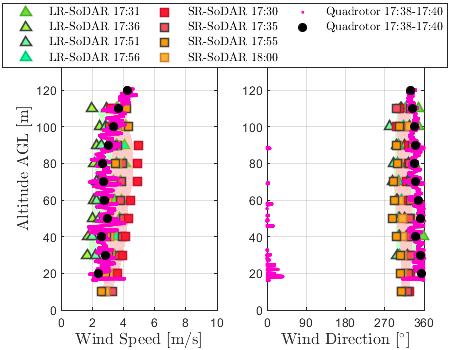}
\label{subfig:1p0ms1}}\\
\subfigure[]{\includegraphics[width = 67mm]{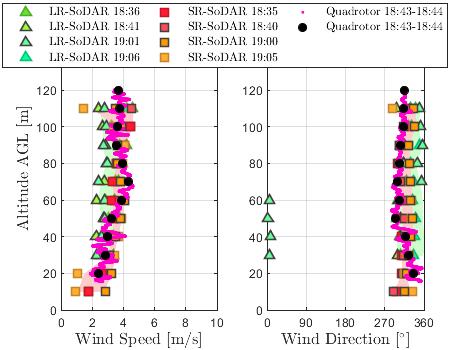}
\label{subfig:1p5ms1}}
\subfigure[]{\includegraphics[width = 67mm]{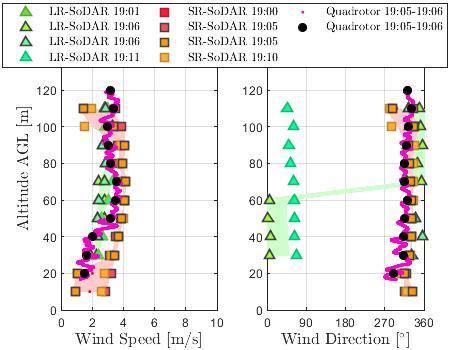}
\label{subfig:1p5ms4}}\\
\caption{ Comparison of quadrotor and SoDAR wind speed and wind direction profile measurements ascending vertically from 10 to 120 m AGL at constant rates of (a) 0.5 m/s, (b) 1 m/s, (c) 1.5 m/s, and (d) 2 m/s.}
\label{fig:wind_valocity_profiles}
\end{figure}

In total, four sets of quadrotor wind velocity profiles corresponding to ascent rates $V_{z_{\rm eq}}>0$ were compared to SoDAR wind observations for validation. Results from validation assessments are shown in Figure~\ref{fig:wind_valocity_profiles} for each ascent rate with quadrotor wind estimates averaged both over 1 second and 10 meter intervals. There was no clear dependency of the performance of the quadrotor wind estimates on ascent rate. Overall, quadrotor wind speed and direction estimates demonstrated good performance tracking SoDAR observations while profiling the lower atmosphere. Especially during the time period that the quadrotor ascended at 1.5 m/s, spatial and temporal variability in wind speed and direction were small and quadrotor winds compared very well with the sodar measurements . All other wind estimates were difficult to corroborate with SoDAR measurements during periods of significant wind variability. These findings are further reinforced by quadrotor and SoDAR comparisons shown in Figures~\ref{fig:ascent_1p0ms}-\ref{fig:ascent_2p0ms} for ascent rates of 1, 1.5 and 2 m/s. 

\section{Discussion}
\label{sec:discussion}
Five nominal models were identified for the control-augmented rigid body dynamics of a quadrotor in both hovering and steady vertical ascending flight. For each model, the  plunge, yaw, roll, and pitch dynamics were characterized separately using step-wise regression and output error methods. Step-wise regression was employed to determine the structure for each model using a set of candidate regressors. Model structures determined from step-wise regression were used to estimate parameter coefficients using the output error method. Trends in parameter variations with respect to ascent rate were found to be monotonic only for a subset of parameters. Significant variations in parameter estimates can be the result of correlation  between regressors used for parameter estimation. Correlation among regressors is likely to occur when regressors change proportionality in system identification experiments. Nonetheless, the identified models were then used to test the observability of wind-augmented models for wind estimation. Future work will focus on improving upon validation experiments to study more closely the sensitivity of model parameters.

Following the identification of quadrotor models, field experiments were conducted to validate quadrotor estimates using wind observations from a sonic anemometer and two SoDARs. Wind observations from ground-based sensors were compared at 10, 30, 70 and 110~m AGL to determine their agreement  prior to validating quadrotor wind estimates. Results from these comparisons show an overall good agreement between sonic anemometer and SR-SoDAR measurements at 10~m AGL and SoDARs at 30, 70, 110~m AGL across the entire time lapse of observations for each sensor pair. The mean bias error and RMSE of observations from the sonic anemometer and SR-SoDAR at 10~m AGL did not exceed values, respectively, 0.7~m/s and 1.0~m/s for wind speed and -0.8$^\circ$ and 19.0$^\circ$ for wind direction. These same error metrics for SoDAR observations at 30, 70, and 110~m AGL had maximum values of -0.7~m/s and 1.2~m/s for wind speed and -7.2$^\circ$  and 34.8$^\circ$ for wind directions. However, acute wind speed and wind direction errors were observed across SoDAR measurements during quadrotor operations. This motivated additional studies to discern the impact of quadrotor operations on the error of SoDAR observations.

To determine the impact of quadrotor operations on the reliability of SoDAR observations, the noise intensity and signal to noise ratio (SNR) of SoDAR wind measurements at heights of 30, 70, and 110~m AGL were assessed during and in the absence of quadrotor flights. Experimental data suggest that quadrotor operations had minimal impact on the accuracy of SoDAR measurements when these devices were measuring wind velocity at 10~m AGL. Quadrotor profiles, on the other hand, were determined to impact SoDAR wind observations between 30 and 120~AGL. This fact is attributed to the quadrotor traversing through the field of view of both SoDARs when sampling at higher altitudes. Because of these observations, the comparisons of quadrotor and SoDAR data were modified to avoid the corrupted SoDAR data.

Quadrotor wind estimates hovering at 10~m AGL were validated first using wind measurements from both the sonic anemometer and the SR-SoDAR. Considering that quadrotor operations at 10 m AGL did not have a significant impact on SoDAR observations, wind velocity measurements were compared to all ground-based assets. Results from the comparison demonstrated good agreement across 5-minute averages in spite of measurements not being sampled at coincident locations. The averaged mean bias error and RMSE values of quadrotor wind speed and wind direction estimates were determined to be 0.3 m/s and 9.9$^\circ$ relative to sonic anemometer observations and 1 m/s and 9.9$^\circ$ relative to SR-SoDAR measurements. In contrast with wind estimation results from ~\cite{gonzalez2019sensing}, comparison results demonstrate an overall improvement in the accuracy of the RBWindPro algorithm estimating wind speed and wind direction using state measurements from a quadrotor hovering.

Validating quadrotor wind profile estimates proved to be difficult for various reasons. First, wind observations from different sensors are inherently not uniform. The non-uniformity across sensors is attributed to how each sensor measures wind velocity, the air volume that is sampled, and the sampling duration. Second, quadrotor operations were found to corrupt SoDAR wind observations. Consequently, quadrotor wind estimates had to be validated using SoDAR observations measured before and after quadrotor operations. These factors limited the performance assessment of quadrotor wind estimation to periods where atmospheric conditions were fairly stationary across the SoDAR sampling domain.

Based on SoDAR observations, wind speed and wind direction were fairly stationary while quadrotor wind profiles were conducted ascending vertically at 1.5 m/s between 18:35 and 19:06 EDT. The quadrotor wind estimates ascending at 1.5 m/s follow wind speed and wind direction SoDAR observations closely. The same is also true for all quadrotor wind profiles conducted during this same period as is shown in Figure~\ref{fig:ascent_1p5ms}. Good performance estimating wind direction was also observed for  quadrotor profiles ascending at 0.5 m/s between 15:30 and 15:46 EDT. All other quadrotor estimates were difficult to validate due to high wind variability during sampling periods. All things considered, positive validation results for a subset of quadrotor wind estimates provides motivation for comprehensive validation experiments that avoid or mitigate for quadrotor and SoDAR interactions. 

Future work will involve improving validation experiments using lessons learned from this study for a more comprehensive performance assessment of quadrotor wind profiling. Improvements to field experiments will require making three modifications. First, one must increase the spatial separation between SoDARs to ensure quadrotor operations do not interfere with wind field measurements. Second, one might incorporate an additional sonic anemometer mounted on a separate quadrotor, to measure wind velocity profiles directly next the quadrotor estimating wind velocity profiles from motion perturbations. Lastly, because of the lateral separation between the ground-based sensors and the quadrotor, one should conduct validation experiments when atmospheric conditions are relatively homogeneous and stationary, and significant uniformity of the wind field sampled by atmospheric sensors is expected. 

\section{Conclusions}
\label{sec:conclusion}

An off-the-shelf quadrotor can be used to obtain model-based wind velocity estimates as long as the motion data logged on board the autopilot is accessible to the user. However, the accuracy of wind velocity estimates depends on how well the motion model characterizes the dynamics of the quadrotor for its operating condition. This paper extends a model based framework exploiting the rigid body dynamics of a quadrotor for hovering-flight wind estimation to estimate wind velocity along a vertical path in the lower atmosphere. The extension involved characterizing rigid body models for equilibrium flight conditions corresponding to each of five steady-ascending rates: $V_{z_{\rm eq}} = \{0.0, 0.5,1.0,1.5,2.0\}$~m/s. Each quadrotor model was characterized employing stepwise regression and output error parameter estimation. An observability analysis confirmed the feasibility of estimating wind velocity using the identified model structures. Trends in parameter estimates also suggest that slower ascent rates may result in more accurate wind estimates. Significant variations in parameter estimates for higher ascent rates can be the outcome of limitations generating manually the rich and precisely timed excitation signals needed for model identification. Further studies are required to investigate this possibility in depth.

Field experiments were conducted to validate quadrotor wind estimates using in-situ and remote-sensing atmospheric sensors. Results from validation experiments demonstrated quadrotor wind estimates in hovering flight to be within within small error of sonic anemometer and SoDAR wind observations. Quadrotor wind profile estimates, on the other hand, were difficult to validate comprehensively because quadrotor operations affect the reliability of SoDAR wind measurements. However, in instances when atmospheric conditions were relatively invariant prior to and after quadrotor operations, quadrotor wind estimates demonstrated  very good agreement with wind speed and wind direction from SoDAR measurements. Overall, this study demonstrates the feasibility of model-based vertical wind profiling using multirotor UAS in the lower atmosphere.

\vspace{6pt} 


\authorcontributions{J.G.R developed the model-based wind sensing methodology presented in this manuscript, characterized vehicle models using aircraft system identification, led field experiments to validate quadrotor wind estimates, curated data from field experiments, and led the writing of the manuscript. S.F.J.D co-led field experiments, provided sonic anemometer wind data, provided guidance for the analysis wind measurements, and assisted in writing the manuscript. S.D.R assisted with validation experiments, provided guidance for the analysis of wind measurements. C.A.W. provided guidance for the analysis of wind measurements, and assisted in writing the manuscript. }

\funding{This research was supported in part by grants from the National Science Foundation (NSF) under grant number AGS 1520825 (Hazards SEES: Advanced Lagrangian Methods for Prediction, Mitigation and Response to Environmental Flow Hazards) and DMS 1821145 (Data-Driven Computation of Lagrangian Transport Structure in Realistic Flows) as well as the NASA Earth and Space Science Fellowship under grant number 80NSSC17K0375. We declare that opinions, findings, and conclusions or recommendations expressed in this material are those of the authors and do not necessarily reflect the views of the sponsors.}

\acknowledgments{We thank Jean-Michel Fahmi, Virginia Tech, for serving as a pilot in command for some UAS missions conducted for media and documentation purposes.}

\conflictsofinterest{The authors declare no conflict of interest.} 

\abbreviations{The following abbreviations are used in this manuscript:\\
\noindent 
\begin{tabular}{@{}ll}
ABL & Atmospheric boundary layer \\
Cov & Covariance \\
LR & Long range \\
LTI & Linear time invariant \\
MBE & Mean bias error\\
RMSE & Root mean squared error\\
SA & Sonic Anemometer \\
SNR & Signal-to-noise ratio \\
SoDAR & Sounding detection and ranging\\
SR & Short range\\
UAS& Small unmanned aircraft systems \\
ctrl & Control\\
ref & Reference\\
\end{tabular}}
\appendixtitles{yes} 
\appendix

\section{Quadrotor Wind Velocity Profiles}
\label{s:quadrotor_wind_velocity_profiles}

\begin{figure}[tbh!]
\centering
\subfigure[]{\includegraphics[width = 67mm]{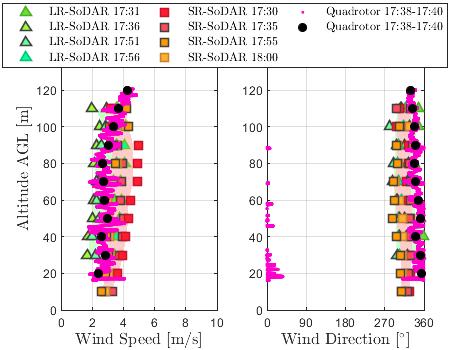}}
\subfigure[]{\includegraphics[width = 67mm]{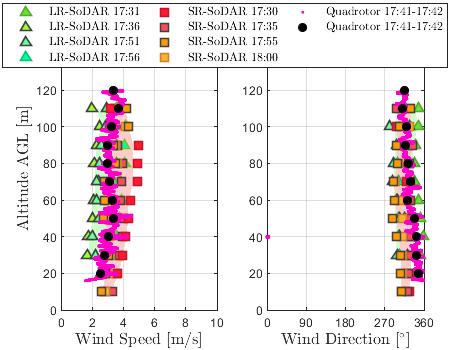}}
\subfigure[]{\includegraphics[width = 67mm]{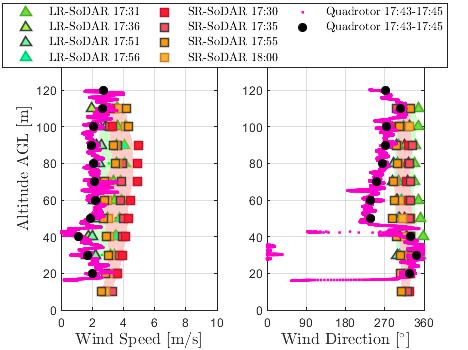}}
\caption{Comparison of quadrotor and SoDAR wind speed and wind direction profile measurements ascending vertically from 10 to 120 m AGL at a constant rate of 1 m/s.}
\label{fig:ascent_1p0ms}
\end{figure}

\begin{figure}[tbh!]
\centering
\subfigure[]{\includegraphics[width = 67mm]{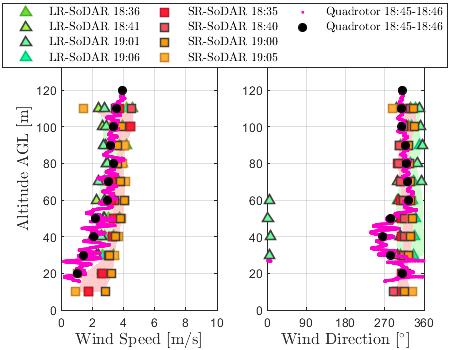}}
\subfigure[]{\includegraphics[width = 67mm]{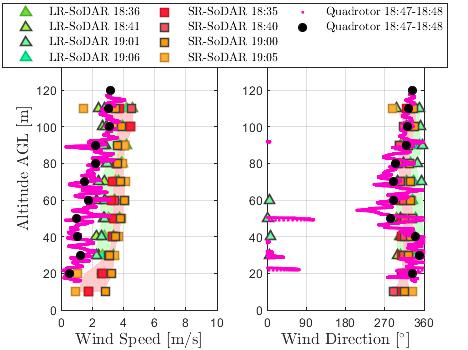}}\\
\subfigure[]{\includegraphics[width = 67mm]{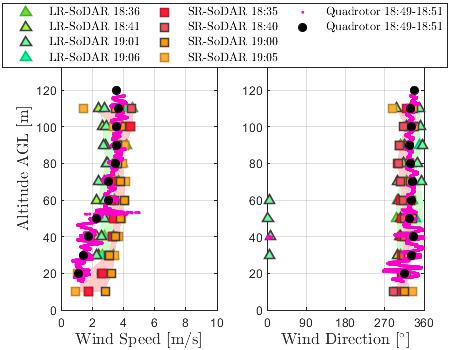}}
\subfigure[]{\includegraphics[width = 67mm]{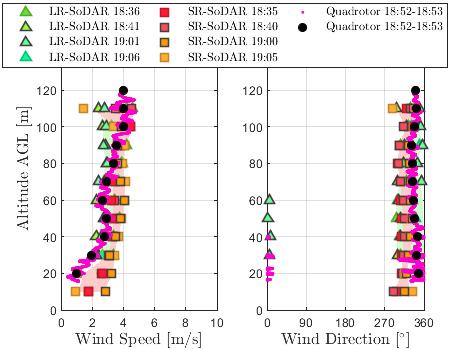}}
\label{subfig:1}
\caption{Comparison of quadrotor and SoDAR wind speed and wind direction profile measurements ascending vertically from 10 to 120 m AGL at a constant rate of 1.5 m/s.}
\label{fig:ascent_1p5ms}
\end{figure}

\begin{figure}[tbh!]
\centering
\subfigure[]{\includegraphics[width = 67mm]{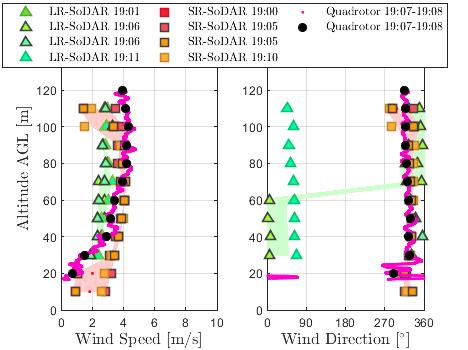}}
\subfigure[]{\includegraphics[width =67mm]{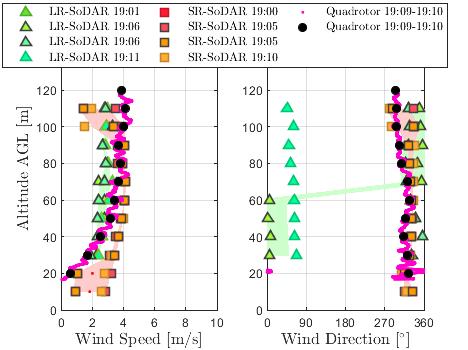}}\\
\subfigure[]{\includegraphics[width = 67mm]{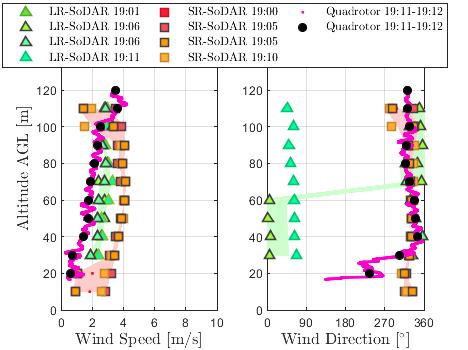}}
\subfigure[]{\includegraphics[width = 67mm]{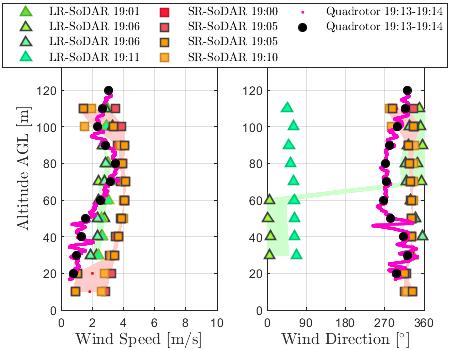}}
\caption{ Comparison of quadrotor and SoDAR wind speed and wind direction profile measurements ascending vertically from 10 to 120 m AGL at a constant rate of 2 m/s.}
\label{fig:ascent_2p0ms}
\end{figure}

Additional quadrotor wind velocity profiles corresponding to constant vertical ascent rates of 1, 1.5 and 2 m/s are shown in Figures~\ref{fig:ascent_1p0ms}-\ref{fig:ascent_2p0ms}. 

\section{A Reliability Study of SoDAR Wind Measurements}
\label{s:a_reliability_study_of_SoDAR_Wind_Measurements}

A two-part reliability study was conducted to investigate anomalies detected in SoDAR wind observations during quadrotor operations. The first part of the study looked at the spatial footprint of quadrotor operations relative to both the position and viewing angle of each SoDAR. The spatial footprint of quadrotor operations relative to SoDAR wind observations was determined from GPS position information provided by the quadrotor's autopilot computer and SoDAR data logs. Quadrotor and SoDAR position information was used to determine if airframe obstruction of acoustic signals or propeller downwash corrupted SoDAR wind measurements. The second part of the study examined both the signal-to-noise-ratio (SNR) and noise intensity corresponding to wind measurements from each SoDAR prior to and during quadrotor operations. Combined, SNR and noise intensity SoDAR measurements were used to determine if anomalies in wind observations were attributed to quadrotor noise during flight operations.  Findings from the two-part study can be used to inform best practices for integrating quadorotor and SoDAR operations for atmospheric wind sensing.

\begin{figure}[tbh!]
    \centering
    \includegraphics[width= 11 cm]{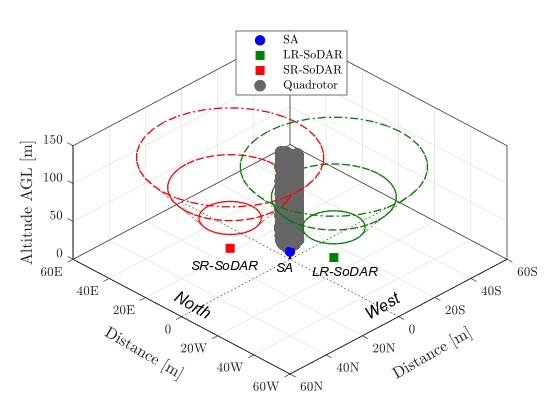}
    \caption{Spatial footprint of quadrotor operations relative to ground-based atmospheric sensors. }
    \label{fig:sensor_location}
\end{figure}

\begin{figure}[tbh!]
\centering
\subfigure[]{\includegraphics[width = 13 cm]{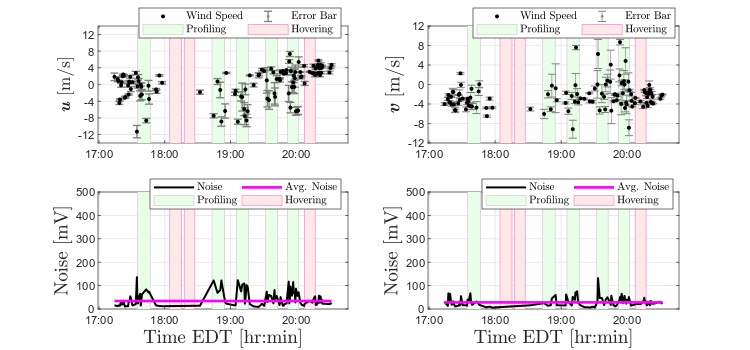}
\label{subfig:SR30}} 
\subfigure[]{\includegraphics[width = 13 cm]{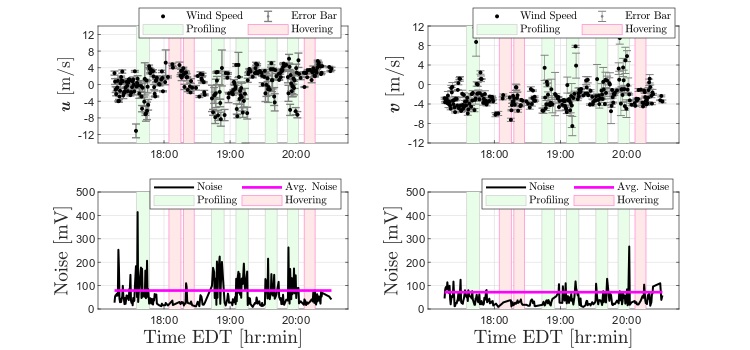}
\label{subfig:SR60}}
\subfigure[]{\includegraphics[width = 13 cm]{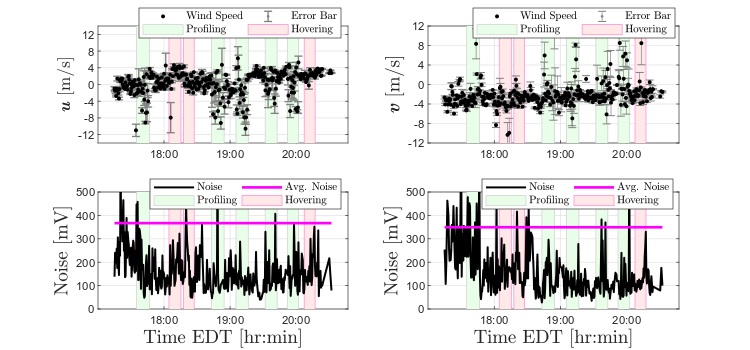}
\label{subfig:SR90}}
\caption{Noise intensity values for $\bm{u}$ and $\bm{v}$ wind velocity observations from the LR-SoDAR at (a) 110 m AGL, (b) 70 m AGL, and (c) 30 m AGL.}

  	\label{fig:LR-SoDAR_SR}
  \end{figure}

\begin{figure}[tbh!]
\centering
\subfigure[]{\includegraphics[width = 13 cm]{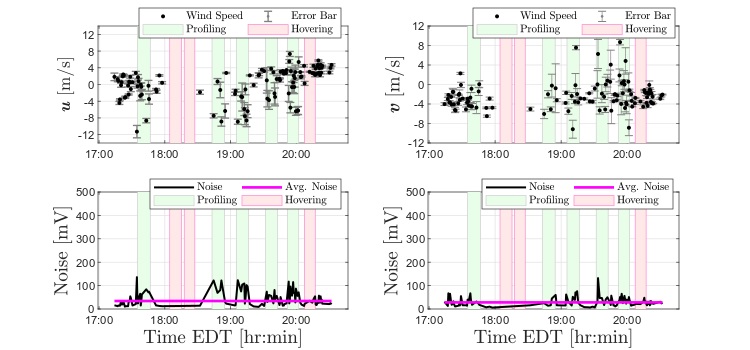}
\label{subfig:SNR30}} 
\subfigure[]{\includegraphics[width = 13 cm]{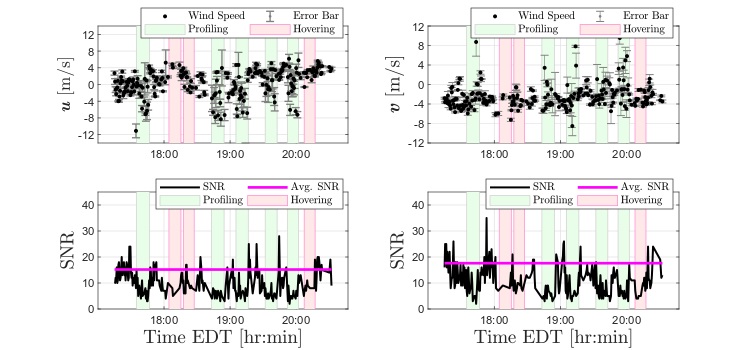}
\label{subfig:SNR60}}
\subfigure[]{\includegraphics[width = 13 cm]{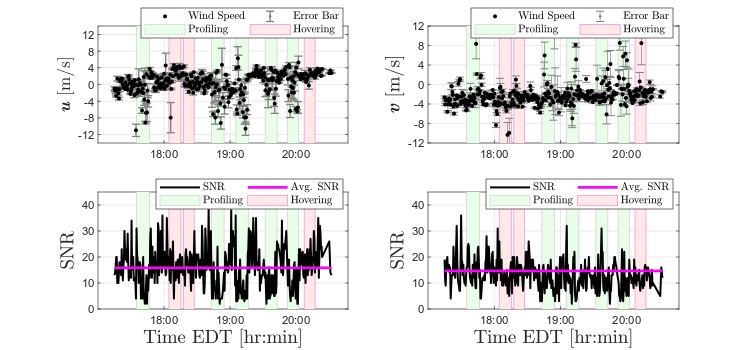}
\label{subfig:SNR90}}
\caption{Singal-to-noise ratios for $\bm{u}$ and $\bm{v}$ wind velocity observations from the LR-SoDAR at (a) 110 m AGL, (b) 70 m AGL, and (c) 30 m AGL.}

  	\label{fig:LR-SoDAR_SNR}
  \end{figure}

\begin{figure}[tbh!]
\centering
\subfigure[]{\includegraphics[width = 13 cm]{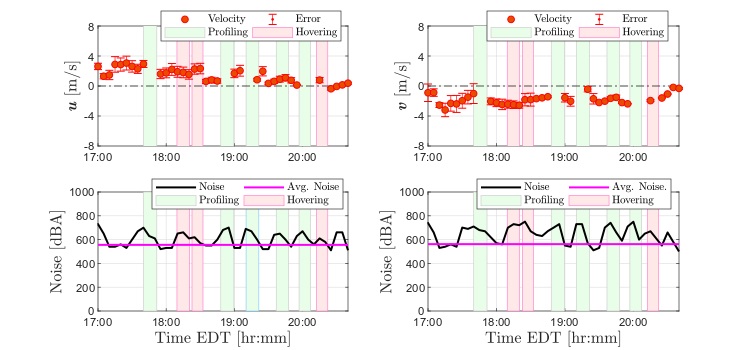}
\label{subfig:PA-0noise110}} \qquad
\subfigure[]{\includegraphics[width = 13 cm]{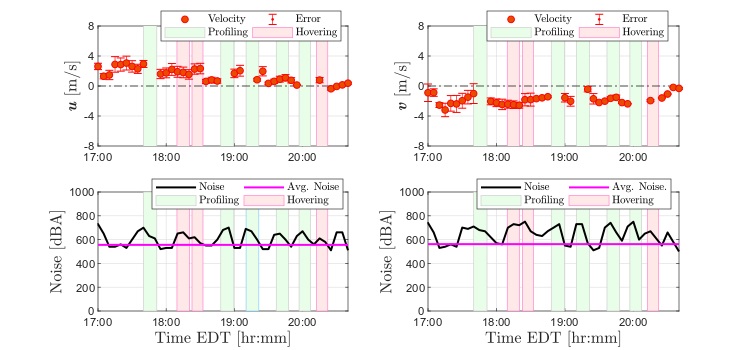}
\label{subfig:PA-0vnoise70}}\\
\subfigure[]{\includegraphics[width = 13 cm]{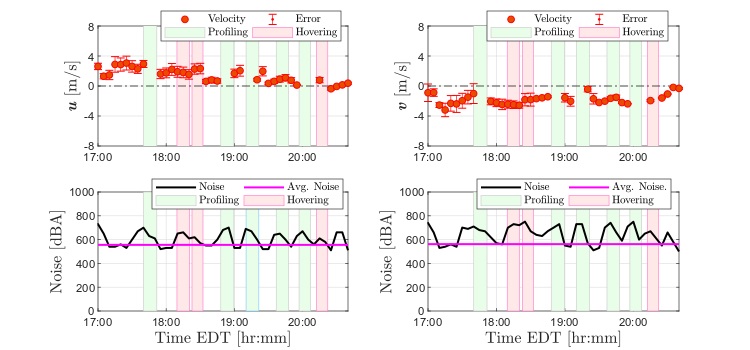}
\label{subfig:PA-0noise30}}
\caption{Noise intensity for $\bm{u}$ and $\bm{v}$ wind velocity observations from the SR-SoDAR at (a) 110 m AGL, (b) 70 m AGL, and (c) 30 m AGL.}
\label{fig:SR_Noise}
\end{figure}

\begin{figure}[tbh!]
\centering
\subfigure[]{\includegraphics[width =13 cm]{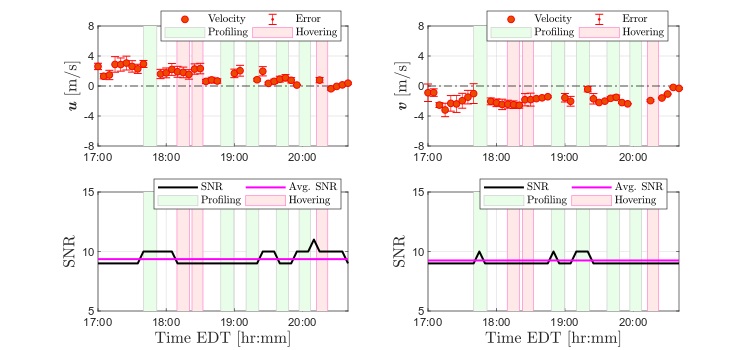}
\label{subfig:PA-0snr110}} \qquad
\subfigure[]{\includegraphics[width = 13 cm]{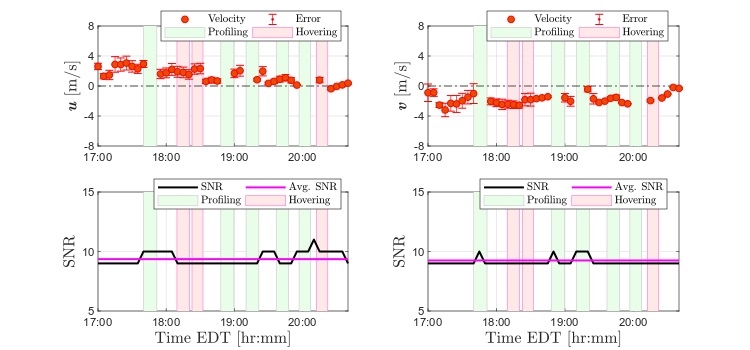}
\label{subfig:PA-0snr70}}\\
\subfigure[]{\includegraphics[width = 13 cm]{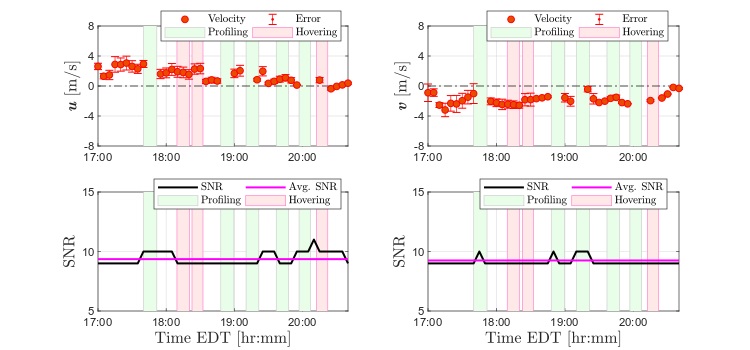}
\label{subfig:PA-0snr30}}
\caption{Signal-to-noise ratios for $\bm{u}$ and $\bm{v}$ wind velocity observations from the SR-SoDAR at (a) 110 m AGL, (b) 70 m AGL, and (c) 30 m AGL.}
\label{fig:SR-SoDAR_SNR}
\end{figure}

 From the two-part reliability study it was determined that quadrotor flight operations impact SoDAR wind observations when operating in close proximity. Assessment of quadrotor's  flight path showed the quadrotor profiling through the sampling volumes of both the LR-SoDAR and SR-SoDAR 60 m AGL during flight operations. A 3-D rendering of this result is shown in Figure~\ref{fig:sensor_location} where the ground position of the quadrotor and two SoDARs are plotted on the plot's x-y plane relative sonic anemometer and the height of measurements is plotted on the z axis. Additionally, sudden changes in noise intensity and SNR during quadrotor operations were observed to coincide with corrupted wind measurements as shown in Figures~\ref{fig:LR-SoDAR_SR}-\ref{fig:SR-SoDAR_SNR}. For example, $\bm{u}$ and $\bm{v}$ wind velocity components measured at 30, 70, and 110 m AGL with the LR-SoDAR show an abrupt increase in magnitude during quadrotor flights. Observations of $\bm{u}$ and $\bm{v}$ wind velocity components at 30, 70, and 110 m AGL were not logged by the SR-SoDAR during quadrotor operations. Both outcomes hinder the validation of quadrotor wind estimates at higher altitudes. 
 
 Results from the two-part study demonstrate a strong relationship between quadrotor noise and anomalies found in SoDAR wind observations. Based on our findings, quadrotor operations can interfere significantly with ground-based acoustic wind measurements. Precaution should be exercised when operating multirotor aircraft near SoDARs. Users will have to gauge a  safe distance of separation based on the sampling volume of the SoDAR and the size of the multirotor aircraft used in flight operations. 
\reftitle{References}

\externalbibliography{yes}

\pagebreak
\bibliography{Biblio1}

\begin{thebibliography}{-------}
\providecommand{\natexlab}[1]{#1}

\bibitem[Gonz\'alez-Rocha \em{et~al.}(2019)Gonz\'alez-Rocha, Woolsey, Sultan,
  and De~Wekker]{gonzalez2019sensing}
Gonz\'alez-Rocha, J.; Woolsey, C.A.; Sultan, C.; De~Wekker, S.F.J.
\newblock Sensing wind from quadrotor motion.
\newblock {\em Journal of Guidance, Control, and Dynamics} {\bf 2019}, {\em
  42},~836--852.

\bibitem[Barbieri \em{et~al.}(2019)Barbieri, Kral, Bailey, Frazier, Jacob,
  Reuder, Brus, Chilson, Crick, Detweiler, Doddi, Elston, Foroutan,
  Gonz\'alez-Rocha, Greene, Guzman, Houston, Islam, Kemppinen, Lawrence,
  Pillar-Little, Ross, Sama, Schmale, Schuyler, Shankar, Smith, Waugh, Dixon,
  Borenstein, and de~Boer]{barbieri2019intercomparison}
Barbieri, L.; Kral, S.T.; Bailey, S.C.C.; Frazier, A.E.; Jacob, J.D.; Reuder,
  J.; Brus, D.; Chilson, P.B.; Crick, C.; Detweiler, C.; Doddi, A.; Elston, J.;
  Foroutan, H.; Gonz\'alez-Rocha, J.; Greene, B.R.; Guzman, M.I.; Houston,
  A.L.; Islam, A.; Kemppinen, O.; Lawrence, D.; Pillar-Little, E.A.; Ross,
  S.D.; Sama, M.P.; Schmale, David~G., I.; Schuyler, T.J.; Shankar, A.; Smith,
  S.W.; Waugh, S.; Dixon, C.; Borenstein, S.; de~Boer, G.
\newblock Intercomparison of small unmanned aircraft system (sUAS) measurements
  for atmospheric science during the LAPSE-RATE campaign.
\newblock {\em Sensors} {\bf 2019}, {\em 19},~2179.

\bibitem[Jacob \em{et~al.}(2018)Jacob, Chilson, Houston, and
  Smith]{jacob2018considerations}
Jacob, J.; Chilson, P.; Houston, A.; Smith, S.
\newblock Considerations for atmospheric measurements with small unmanned
  aircraft systems.
\newblock {\em Atmosphere} {\bf 2018}, {\em 9},~252.

\bibitem[Chilson \em{et~al.}(2019)Chilson, Bell, Brewster, de~Azevedo, Carr,
  Carson, Doyle, Fiebrich, Greene, Grimsley, Kanneganti, Martin, Moore, Palmer,
  Pillar-Little, Salazar-Cerreno, Segales, Weber, Yeary, and
  Droegemeier]{chilson2019moving}
Chilson, P.B.; Bell, T.M.; Brewster, K.A.; de~Azevedo, G.B.H.; Carr, F.H.;
  Carson, K.; Doyle, W.; Fiebrich, C.A.; Greene, B.R.; Grimsley, J.L.;
  Kanneganti, S.T.; Martin, J.; Moore, A.; Palmer, R.D.; Pillar-Little, E.A.;
  Salazar-Cerreno, J.L.; Segales, A.R.; Weber, M.E.; Yeary, M.; Droegemeier,
  K.K.
\newblock Moving towards a network of autonomous UAS atmospheric profiling
  stations for observations in the Earth’s lower atmosphere: The 3D mesonet
  oncept.
\newblock {\em Sensors} {\bf 2019}, {\em 19},~2720.

\bibitem[Smith \em{et~al.}(2017)Smith, Chilson, Houston, and
  Jacob]{smith2017catalyzing}
Smith, S.W.; Chilson, P.B.; Houston, A.L.; Jacob, J.D.
\newblock Catalyzing collaboration for multi-disciplinary UAS development with
  a flight campaign focused on meteorology and atmospheric physics.
\newblock  AIAA Information Systems-AIAA Infotech@ Aerospace,  2017, p. 1156.

\bibitem[Villa \em{et~al.}(2016)Villa, Gonzalez, Miljievic, Ristovski, and
  Morawska]{villa2016overview}
Villa, T.; Gonzalez, F.; Miljievic, B.; Ristovski, Z.; Morawska, L.
\newblock An overview of small unmanned aerial vehicles for air quality
  measurements: Present applications and future prospectives.
\newblock {\em Sensors} {\bf 2016}, {\em 16},~1072.

\bibitem[Nolan \em{et~al.}(2018)Nolan, Pinto, Gonz{\'a}lez-Rocha, Jensen,
  Vezzi, Bailey, de~Boer, Diehl, Laurence, Powers, Ross, and
  Schmale~III]{nolan2018coordinated}
Nolan, P.J.; Pinto, J.; Gonz{\'a}lez-Rocha, J.; Jensen, A.; Vezzi, C.; Bailey,
  S.; de~Boer, G.; Diehl, C.; Laurence, R.; Powers, C.; Ross, S.D.;
  Schmale~III, D.G.
\newblock Coordinated unmanned aircraft system ({UAS}) and ground-based weather
  measurements to predict {L}agrangian coherent structures ({LCS}s).
\newblock {\em Sensors} {\bf 2018}, {\em 18},~4448.

\bibitem[Nolan \em{et~al.}(2019)Nolan, McClelland, Woolsey, and
  Ross]{nolan2019method}
Nolan, P.J.; McClelland, H.G.; Woolsey, C.A.; Ross, S.D.
\newblock A method for detecting atmospheric {L}agrangian coherent structures
  using a single fixed-wind unmanned aircraft system.
\newblock {\em Sensors} {\bf 2019}, {\em 19},~1607.

\bibitem[Carranza \em{et~al.}(2018)Carranza, Rafiq, Frausto-Vicencio, Hopkins,
  Verhulst, Rao, Duren, and Miller]{carranza2018vista}
Carranza, V.; Rafiq, T.; Frausto-Vicencio, I.; Hopkins, F.M.; Verhulst, K.R.;
  Rao, P.; Duren, R.M.; Miller, C.E.
\newblock Vista-LA: Mapping methane-emitting infrastructure in the Los Angeles
  megacity.
\newblock {\em Earth System Science Data} {\bf 2018}, {\em 10},~653.

\bibitem[Chao and Chen(2010)]{chao2010surface}
Chao, H.; Chen, Y.
\newblock Surface wind profile measurement using multiple small unmanned aerial
  vehicles.
\newblock  Proceedings of the 2010 American Control Conference. IEEE,  2010,
  pp. 4133--4138.

\bibitem[Fairley(2018)]{Fairley2018building}
Fairley, P.
\newblock Building a Weather-Smart Grid.
\newblock {\em Scientific American} {\bf 2018}, {\em 319},~60 -- 65.

\bibitem[Phuangpornpitak and Tia(2013)]{phuangpornpitak2013opportunities}
Phuangpornpitak, N.; Tia, S.
\newblock Opportunities and challenges of integrating renewable energy in smart
  grid system.
\newblock {\em Energy Procedia} {\bf 2013}, {\em 34},~282--290.

\bibitem[Colak \em{et~al.}(2015)Colak, Fulli, Bayhan, Chondrogiannis, and
  Demirbas]{colak2015critical}
Colak, I.; Fulli, G.; Bayhan, S.; Chondrogiannis, S.; Demirbas, S.
\newblock Critical aspects of wind energy systems in smart grid applications.
\newblock {\em Renewable and Sustainable Energy Reviews} {\bf 2015}, {\em
  52},~155--171.

\bibitem[Wildmann \em{et~al.}(2017)Wildmann, Bernard, and
  Bange]{wildmann2017measuring}
Wildmann, N.; Bernard, S.; Bange, J.
\newblock Measuring the local wind field at an escarpment using small
  remotely-piloted aircraft.
\newblock {\em Renewable energy} {\bf 2017}, {\em 103},~613--619.

\bibitem[Alsalous \em{et~al.}(2017)Alsalous, Galaviz, and
  Gulding]{alsalous2017evaluation}
Alsalous, O.; Galaviz, R.; Gulding, J.
\newblock Evaluation of the efficiency of traffic management initiatives wind
  delays.
\newblock  17th AIAA Aviation Technology, Integration, and Operations
  Conference,  2017, p. 4263.

\bibitem[Tang \em{et~al.}(2010)Tang, Chan, and Haller]{tang2010accurate}
Tang, W.; Chan, P.W.; Haller, G.
\newblock Accurate extraction of Lagrangian coherent structures over finite
  domains with application to flight data analysis over Hong Kong International
  Airport.
\newblock {\em Chaos: An Interdisciplinary Journal of Nonlinear Science} {\bf
  2010}, {\em 20},~017502.

\bibitem[Tang \em{et~al.}(2011)Tang, Chan, and Haller]{tang2011lagrangian}
Tang, W.; Chan, P.W.; Haller, G.
\newblock Lagrangian coherent structure analysis of terminal winds detected by
  lidar. Part I: Turbulence structures.
\newblock {\em Journal of Applied Meteorology and Climatology} {\bf 2011}, {\em
  50},~325--338.

\bibitem[Knutson \em{et~al.}(2015)Knutson, Tang, and
  Chan]{knutson2015lagrangian}
Knutson, B.; Tang, W.; Chan, P.W.
\newblock Lagrangian coherent structure analysis of terminal winds:
  Three-dimensionality, intramodel variations, and flight analyses.
\newblock {\em Advances in Meteorology} {\bf 2015}, {\em 2015}.

\bibitem[Rabinovich \em{et~al.}(2018)Rabinovich, Curry, and
  Elkaim]{rabinovich2018toward}
Rabinovich, S.; Curry, R.E.; Elkaim, G.H.
\newblock Toward dynamic monitoring and suppressing uncertainty in wildfire by
  multiple unmanned air vehicle system.
\newblock {\em Journal of Robotics} {\bf 2018}, pp. 1 -- 12.

\bibitem[da~Silva \em{et~al.}(2017)da~Silva, Bernardo, de~Oliveira, and
  Rosa]{da2017unmanned}
da~Silva, L.C.B.; Bernardo, R.M.; de~Oliveira, H.A.; Rosa, P.F.F.
\newblock Unmanned aircraft system coordination for persistent surveillance
  with different priorities.
\newblock  2017 IEEE 26th International Symposium on Industrial Electronics
  (ISIE). IEEE,  2017, pp. 1153--1158.

\bibitem[AL-Dhief \em{et~al.}(2017)AL-Dhief, Sabri, Fouad, Latiff, and
  Albader]{al2017review}
AL-Dhief, F.T.; Sabri, N.; Fouad, S.; Latiff, N.A.; Albader, M.A.A.
\newblock A review of forest fire surveillance technologies: Mobile ad-hoc
  network routing protocols perspective.
\newblock {\em Journal of King Saud University-Computer and Information
  Sciences} {\bf 2017}.

\bibitem[{Xing} \em{et~al.}(2019){Xing}, {Zhang}, {Su}, {Qu}, and
  {Yu}]{Xingetal2019}
{Xing}, Z.; {Zhang}, Y.; {Su}, C.; {Qu}, Y.; {Yu}, Z.
\newblock Kalman filter-based wind estimation for forest fire monitoring with a
  quadrotor UAV.
\newblock  2019 IEEE Conference on Control Technology and Applications (CCTA),
  2019, pp. 783--788.

\bibitem[Duren \em{et~al.}(2019)Duren, Thorpe, Foster, Rafiq, Hopkins, Yadav,
  Bue, Thompson, Conley, Colombi, et~al.]{duren2019california}
Duren, R.M.; Thorpe, A.K.; Foster, K.T.; Rafiq, T.; Hopkins, F.M.; Yadav, V.;
  Bue, B.D.; Thompson, D.R.; Conley, S.; Colombi, N.K.; others.
\newblock California’s methane super-emitters.
\newblock {\em Nature} {\bf 2019}, {\em 575},~180--184.

\bibitem[{Smith} \em{et~al.}(2017){Smith}, {John}, {Christensen}, and
  {Chen}]{smith2017fugitive}
{Smith}, B.J.; {John}, G.; {Christensen}, L.E.; {Chen}, Y.
\newblock Fugitive methane leak detection using sUAS and miniature laser
  spectrometer payload: System, application and groundtruthing tests.
\newblock  2017 International Conference on Unmanned Aircraft Systems (ICUAS),
  2017, pp. 369--374.

\bibitem[Andersen \em{et~al.}(2018)Andersen, Scheeren, Peters, and
  Chen]{andersen2018auav}
Andersen, T.; Scheeren, B.; Peters, W.; Chen, H.
\newblock A UAV-based active AirCore system for measurements of greenhouse
  gases.
\newblock {\em Atmospheric Measurement Techniques} {\bf 2018}, {\em
  11},~2683--2699.

\bibitem[Rold{\'a}n \em{et~al.}(2015)Rold{\'a}n, Joossen, Sanz, del Cerro, and
  Barrientos]{roldan2015mini}
Rold{\'a}n, J.; Joossen, G.; Sanz, D.; del Cerro, J.; Barrientos, A.
\newblock Mini-UAV based sensory system for measuring environmental variables
  in greenhouses.
\newblock {\em Sensors} {\bf 2015}, {\em 15},~3334--3350.

\bibitem[Greene \em{et~al.}(2019)Greene, Segales, Bell, Pillar-Little, and
  Chilson]{greene2019environmental}
Greene, B.R.; Segales, A.R.; Bell, T.M.; Pillar-Little, E.A.; Chilson, P.B.
\newblock Environmental and sensor integration influences on temperature
  measurements by rotary-wing unmanned aircraft systems.
\newblock {\em Sensors} {\bf 2019}, {\em 19},~1470.

\bibitem[Varentsov \em{et~al.}(2019)Varentsov, Artamonov, Pashkin, and
  Repina]{varentsov2019experience}
Varentsov, M.; Artamonov, A.Y.; Pashkin, A.; Repina, I.
\newblock Experience in the quadcopter-based meteorological observations in the
  atmospheric boundary layer.
\newblock  IOP Conference Series: Earth and Environmental Science. IOP
  Publishing,  2019, p. 012053.

\bibitem[Wolf \em{et~al.}(2017)Wolf, Hardis, Woodrum, Galan, Wichelt, Metzger,
  Bezzo, Lewin, and de~Wekker]{wolf2017wind}
Wolf, C.A.; Hardis, R.P.; Woodrum, S.D.; Galan, R.S.; Wichelt, H.S.; Metzger,
  M.C.; Bezzo, N.; Lewin, G.C.; de~Wekker, S.F.
\newblock Wind Data Collection Techniques on a Multi-rotor Platform.
\newblock  Systems and Information Engineering Design Symposium (SIEDS), 2017.
  IEEE,  2017, pp. 32--37.

\bibitem[de~Boisblanc \em{et~al.}(2014)de~Boisblanc, Dodbele, Kussmann,
  Mukherji, Chestnut, Phelps, Lewin, and de~Wekker]{de2014designing}
de~Boisblanc, I.; Dodbele, N.; Kussmann, L.; Mukherji, R.; Chestnut, D.;
  Phelps, S.; Lewin, G.C.; de~Wekker, S.
\newblock Designing a hexacopter for the collection of atmospheric flow data.
\newblock  Systems and Information Engineering Design Symposium (SIEDS), 2014.
  IEEE,  2014, pp. 147--152.

\bibitem[Donnell \em{et~al.}(2018)Donnell, Feight, Lannan, and
  Jacob]{donnell2018wind}
Donnell, G.W.; Feight, J.A.; Lannan, N.; Jacob, J.D.
\newblock Wind characterization using onboard IMU of sUAS.
\newblock  2018 Atmospheric Flight Mechanics Conference,  2018, p. 2986.

\bibitem[Hollenbeck \em{et~al.}(2018)Hollenbeck, Nunez, Christensen, and
  Chen]{hollenbeck2018wind}
Hollenbeck, D.; Nunez, G.; Christensen, L.E.; Chen, Y.
\newblock Wind measurement and estimation with small unmanned aerial systems
  (suas) using on-board mini ultrasonic anemometers.
\newblock  2018 International Conference on Unmanned Aircraft Systems (ICUAS).
  IEEE,  2018, pp. 285--292.

\bibitem[{Hollenbeck} \em{et~al.}(2019){Hollenbeck}, {Oyama}, {Garcia}, and
  {Chen}]{Hollerbeck2019pitch}
{Hollenbeck}, D.; {Oyama}, M.; {Garcia}, A.; {Chen}, Y.
\newblock Pitch and roll effects of on-board wind measurements using sUAS.
\newblock  2019 International Conference on Unmanned Aircraft Systems (ICUAS),
  2019, pp. 1249--1254.

\bibitem[Prudden \em{et~al.}(2016)Prudden, Fisher, Mohamed, and
  Watkins]{prudden2016flying}
Prudden, S.; Fisher, A.; Mohamed, A.; Watkins, S.
\newblock A flying anemometer quadrotor: Part 1.
\newblock  Proceedings of the 7th International Micro Air Vehicle Conference
  and Competition - Past, Present and Future. IMAV,  2016, pp. 15--21.

\bibitem[Neumann and Bartholmai(2015)]{neumann2015real}
Neumann, P.P.; Bartholmai, M.
\newblock Real-time wind estimation on a micro unmanned aerial vehicle using
  its inertial measurement unit.
\newblock {\em Sensors and Actuators A: Physical} {\bf 2015}, {\em
  235},~300--310.

\bibitem[Brosy \em{et~al.}(2017)Brosy, Krampf, Zeeman, Wolf, Junkermann,
  Sch{\"a}fer, Emeis, and Kunstmann]{brosy2017simultaneous}
Brosy, C.; Krampf, K.; Zeeman, M.; Wolf, B.; Junkermann, W.; Sch{\"a}fer, K.;
  Emeis, S.; Kunstmann, H.
\newblock Simultaneous multicopter-based air sampling and sensing of
  meteorological variables.
\newblock {\em Atmospheric Measurement Techniques} {\bf 2017}, {\em 10},~2773.

\bibitem[Palomaki \em{et~al.}(2017)Palomaki, Rose, van~den Bossche, Sherman,
  and De~Wekker]{palomaki2017wind}
Palomaki, R.T.; Rose, N.T.; van~den Bossche, M.; Sherman, T.J.; De~Wekker, S.F.
\newblock Wind estimation in the lower atmosphere using multirotor aircraft.
\newblock {\em Journal of Atmospheric and Oceanic Technology} {\bf 2017}, {\em
  34},~1183--1191.

\bibitem[Gonzalez-Rocha \em{et~al.}(2017)Gonzalez-Rocha, Woolsey, Sultan,
  de~Wekker, and Rose]{gonzalez2017measuring}
Gonzalez-Rocha, J.; Woolsey, C.A.; Sultan, C.; de~Wekker, S.; Rose, N.
\newblock Measuring atmospheric winds from quadrotor motion.
\newblock  AIAA Atmospheric Flight Mechanics Conference,  2017, p. 1189.

\bibitem[Klein and Morelli(2006)]{klein2006aircraft}
Klein, V.; Morelli, E.A.
\newblock {\em Aircraft System Identification: Theory and Practice}; American
  Institute of Aeronautics and Astronautics Reston, Va, USA,  2006.

\end{thebibliography}

\end{document}